\PassOptionsToPackage{table}{xcolor}
\documentclass[sigconf,nonacm]{acmart}

\usepackage{booktabs}
\usepackage{listings}
\usepackage{url}
\usepackage{tabularx}
\usepackage{graphicx}
\usepackage{array}
\usepackage{colortbl}
\usepackage[most]{tcolorbox}
\usepackage{enumitem}

\setlist[description]{style=unboxed, leftmargin=0pt, labelsep=0.4em,
                      topsep=3pt, partopsep=0pt, itemsep=3pt, parsep=0pt,
                      font=\normalfont\bfseries}

\hypersetup{hidelinks}

\renewcommand\footnotetextcopyrightpermission[1]{}
\makeatletter

\newcommand{\cmdd}[1]{\textcolor{red}{#1}}
\newcommand{\attackfontsize}{\fontsize{9}{11}\selectfont}

\AtBeginDocument{%
}

\title{Beyond the Mandate: A Systematic Security Analysis of the Agent Payments Protocol (AP2)}

\author{Avital Aviv}
\authornote{Both authors contributed equally to this research.}
\email{avitalos6@gmail.com}
\affiliation{%
  \institution{Ben-Gurion University of the Negev}
  \city{Beer-Sheva}
  \country{Israel}
}

\author{Parth A. Gandh}
\authornotemark[1]
\email{gandhip@post.bgu.ac.il}
\affiliation{%
  \institution{Ben-Gurion University of the Negev}
  \city{Beer-Sheva}
  \country{Israel}
}

\author{Ron Bitton}
\email{ron\_bitton@intuit.com}
\affiliation{%
  \institution{Intuit}
  \city{Petah Tikva}
  \country{Israel}
}

\author{Asaf Shabtai}
\email{shabtaia@bgu.ac.il}
\affiliation{%
  \institution{Ben-Gurion University of the Negev}
  \city{Beer-Sheva}
  \country{Israel}
}

\renewcommand{\shortauthors}{Aviv et al.}

\begin{document}

\begin{abstract}
The Agent Payments Protocol (AP2), introduced by Google, enables large language model (LLM)-driven shopping agents to authorize and execute payments on behalf of users. Its signed Checkout and Payment Mandates protect the integrity of transaction data after signing. Agent interactions and external inputs that shape a transaction before authorization remain outside that protection, including Agent-to-Agent Protocol (A2A) messages and Model Context Protocol (MCP) tool calls. Prior work identified replay and prompt-injection attacks in AP2~v0.1. AP2~v0.2 addresses some of these issues but adds capabilities and deployment assumptions that require renewed analysis. We present a systematic security analysis of AP2~v0.2 based on its roles, transaction lifecycle, deployment architectures, and trust boundaries. We divide the lifecycle into five phases and identify five deployment architectures. Using MAESTRO (Multi-Agent Environment, Security, Threat, Risk, Outcome), we model four threat actors, eleven attack surfaces, eighteen adversary capabilities, and six attacker goals. The resulting catalog contains 48 threats spanning five attack families. We score these threats with the Artificial Intelligence Vulnerability Scoring System (AIVSS), identifying eight that reach the High band in at least one architecture. Because no complete public AP2 deployment was available, we build a testbed spanning all five architectures and develop five proof-of-concept demonstrations covering all eight High-risk threats and their mitigations. We also develop a deployment-aware scanner that maps applicable threats to static, cross-role consistency, and adversarial checks. Our analysis shows that valid mandate signatures alone do not ensure that an agent-mediated transaction reflects the user's intent when its pre-authorization context is manipulated.
\end{abstract}

\maketitle

\section{Introduction}
\label{sec1}

Large language model (LLM)-driven agents are beginning to act on users' behalf in the financial sector to browse catalogs, assemble carts and execute payments without the need for a human to approve each step. Google published the Agent Payments Protocol (AP2) to provide a secure framework for agent-led payments~\cite{ap2spec}. AP2~v0.1, released in 2025, introduced the protocol's roles and a chain of signed mandates built on Verifiable Digital Credentials (VDCs), focusing mainly on Human-Present (HP) flows in which the user remains in session and explicitly authorizes each closed mandate through a Trusted Surface~\cite{ap2spec}. AP2~v0.2, released in April 2026, adds Human-Not-Present (HNP) flows for autonomous transactions and defenses against replay attacks~\cite{ap2spec}.

Because AP2 uses LLM-driven agents in protocol roles, it faces threats that classical payment protocols never encounter, such as prompt injection, tool-result poisoning and agent manipulation threats. The specification treats all LLM-driven agentic roles as potential attackers~\cite{ap2spec}. 
AP2 mainly secures the signed mandates and receipts; the pre-signing context that shapes them, including catalog data, tool results and A2A messages, remains outside the signed mandates. Since AP2 also leaves deployment choices to implementers, including whether Model Context Protocol (MCP) is used and whether tool infrastructure is shared, the practical threat surface depends on the architecture. Consequently, an attacker may not need to forge a mandate or receipt to cause harm; instead, the attacker may manipulate the agent’s context, tools or communications so that a validly signed mandate encodes an unsafe or unintended action. These risks connect AP2’s agentic attack surface with its protocol-level security assumptions and show the need for a threat model that covers both signed artifacts and the unsigned context that produces them.

Prior research has addressed individual components of AP2 security, but none provides a systematic threat model for AP2. MCP security research has identified tool poisoning and shadowing attacks~\cite{hou_2025, narajala_habler_2025}, and Louck et al.\ examined the security risks in the A2A protocol~\cite{louck_2025}, yet none of these studies examined whether these threats propagate into AP2 itself. Two other studies examine AP2~v0.1 directly, red-teaming prompt injection in an AP2 shopping agent~\cite{debi_zhu_2026} and characterizing replay and context-binding failures~\cite{lan_2026}, while Mao et al.\ systematize attack vectors across seven agentic commerce protocols including AP2~\cite{mao_2026}. None of these studies analyzed threats across AP2's phases, trust boundaries, and deployment architectures. This paper presents the first threat model that spans AP2~v0.2's phases, trust boundaries, and deployment architectures, validating the high-risk threats on a testbed and providing a security scanner for implementers.

We map AP2's protocol design to concrete security risks. We segment the AP2 transaction lifecycle
into five analytical phases and identify five deployment architectures, then use
MAESTRO~\cite{maestro2025} to enumerate threats across actors, attack surfaces, capabilities, and
goals. We subsequently map each threat to the architectures. This
yields 48 threats grouped five architectures and attack families and assessed with the AIVSS~\cite{aivss} risk-scoring mechanism.

Because no public AP2 deployment existed at the time of our evaluation, we built a custom testbed and developed proof-of-concept attacks against all high-risk threats. This paper presents five of these attacks in detail. A single attack scenario can chain several threats, so the five demonstrations together cover all eight high-risk threats and their mitigations.

We also develop a security scanner that derives a deployment profile, executes the applicable static, cross-role, and adversarial checks, and reports detected AP2 threats. We evaluate the scanner on the same controlled testbed. Our evaluation covers the threat catalog, the AIVSS assessment, and the scanner. A structured STRIDE-GPT comparison tests the catalog's coverage against a separate threat-modeling method. An eight-rater study tests whether independent experts reproduce the derived AIVSS severity bands and finally we perform a layer-ablation study of the scanner. 

In summary, this paper makes five contributions.
\textbf{(1)}~We segment the AP2 transaction lifecycle into five analytical phases and identify five deployment architecture classes, treating a deployment as a distinct class when it exposes at least one threat not represented by an already-defined class.
\textbf{(2)}~We construct the first phase- and architecture-spanning threat model for AP2 using the MAESTRO seven-layer framework, spanning four threat actors, eleven attack surfaces, fourteen access capabilities in four families, four knowledge capabilities, six attacker goals, and the five architectures.
\textbf{(3)}~We catalog 48 AP2 threats and organize them into five attack families based on the AP2 security object each family corrupts, and apply AIVSS separately to each applicable architecture, identifying 8 of the 48 threats as high risk in at least one architecture.
\textbf{(4)}~We develop PoC attacks against all high-risk threats, demonstrating that these threats are realizable along with presenting mitigations for the same.
\textbf{(5)}~We implement a security scanner that profiles AP2 deployments, selects applicable threat checks, and reports detected findings, and evaluate it on our controlled AP2 testbed.

\section{Related Work}
\label{sec}

\subsection{AP2 and Agentic Commerce Security}

Prior work identifies broad risks in autonomous AI systems~\cite{owasp-aai,owasp-llm}. In AP2, these risks threaten mandate integrity, user authorization, and transaction accountability~\cite{ap2spec}.

Two studies analyzed AP2 v0.1. Debi et al.\ showed that prompt injection through malicious content in an AP2-style shopping agent can manipulate rankings and leak user data despite signed mandates~\cite{debi_zhu_2026}. At the execution layer, Lan et al.\ identified replay and context-binding flaws that allow mandates to be reused or rebound without detection~\cite{lan_2026}. These studies expose failures both above and below the cryptographic layer.

Other work examines AP2 within agentic commerce more broadly. Acharya investigated trustless agent payments using decentralized identifiers, verifiable credentials, and zero-knowledge proofs, concluding that protocol-level trust is possible without a central payment intermediary~\cite{acharya_2025}. Mao et al.\ mapped recurring attack classes across seven agentic commerce protocols, including AP2~\cite{mao_2026}. Hu and Rong found that no protocol fully ensures both authorization integrity and settlement accountability~\cite{hu_rong_2025}.

\noindent However, the newly released AP2 v0.2 remains largely unexplored. Existing work does not systematically identify its threats, affected deployment architectures, or architecture-dependent impacts.

\subsection{Security of Supporting Agent Protocols}

AP2 relies on A2A for inter-agent messaging and discovery and on MCP for tool invocation. Because agents consume information from both protocols during transactions, vulnerabilities in either can propagate into AP2.

MCP research identifies tool poisoning, tool shadowing, rug pulls, weak capability attestation, sampling risks, and client exposure to malicious tool metadata~\cite{hou_2025,narajala_habler_2025,jamshidi_2025,maloyan_namiot_2026,huang_2026}. Louck et al.\ analyzed authentication, authorization, integrity, confidentiality, and availability in A2A, ACP, and CORAL~\cite{louck_2025}. However, no study has examined how these threats propagate into AP2 or provided a comprehensive AP2 threat model supported by proof-of-concept demonstrations and mitigations.

This paper addresses that gap and consolidates its findings into a security scanner for AP2 implementers.
\section{AP2 Transaction Lifecycle \& Architectures}

\subsection{AP2 Transaction Lifecycle}
\label{sec:lifecycle}
AP2 secures AI-agent payments with non-repudiable cryptographic proof of user authorization. Its Human-Present (HP) and Human-Not-Present (HNP) modes differ in when authorization occurs and who signs the mandates. This section presents the roles, flows, and five analytical lifecycle phases.

\subsubsection{Roles}\label{subsec:roles}

AP2 specifies five protocol roles~\cite{ap2spec}:
\begin{itemize}
    \item \textbf{Shopping Agent (SA).} Performs product discovery,
    cart assembly, mandate construction, and payment execution on
    behalf of the user.
    \item \textbf{Merchant (M).} Provides the merchant-signed checkout (\texttt{checkout\_jwt}), verifies the Checkout Mandate, and completes the checkout. When the merchant side is implemented by an LLM-driven agent, we refer to it as the Merchant Agent (MA).
    \item \textbf{Credential Provider (CP).} Verifies the Payment
    Mandate and issues a scoped payment credential.
    \item \textbf{Merchant Payment Processor (MPP).} Verifies that the
    payment credential is appropriately scoped to the checkout,
    processes settlement, and returns a Payment Receipt.
    \item \textbf{Trusted Surface (TS).} Renders mandate content to the user, obtains informed consent, and initiates mandate signing.
\end{itemize}

\begin{figure}[t]
    \centering
    \includegraphics[width=1.0\columnwidth]{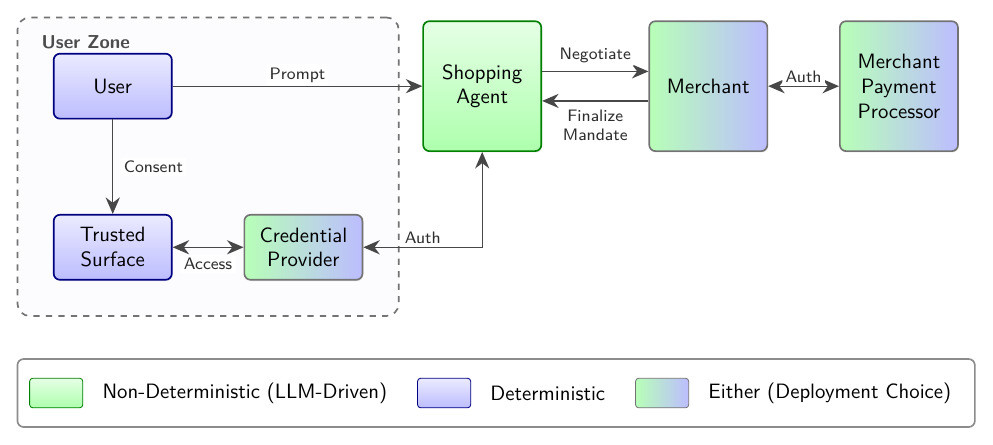}
    \caption{The five AP2 protocol roles and the flow of mandates and receipts among them.}
    \label{fig:roles}
\end{figure}

Figure~\ref{fig:roles} maps their interactions. A role is \emph{agentic} when a non-deterministic LLM handles communication to or from it, making it a potential compromise point that requires tamper-evident defenses. We exclude payment-network and issuer processing, which AP2 does not define or change.

\subsubsection{Mandates}

Mandates are cryptographically signed, tamper-evident AP2 authorization artifacts implemented using Selective Disclosure for JSON Web Tokens (SD-JWTs). AP2 defines Checkout and Payment Mandates, each with open and closed forms~\cite{ap2spec}.

The \textbf{Checkout Mandate} proves to the Merchant that the user authorized a purchase. Its open form captures the user's constraints and goals before a cart is finalized and is used in HNP mode. Its closed form authorizes a specific finalized checkout and includes a \texttt{checkout\_hash} that binds it permanently to that cart.

The \textbf{Payment Mandate} proves to the CP, MPP, and any involved card networks that the user authorized payment for a checkout. Its open form records payment constraints for HNP execution; its closed form authorizes a specific amount. The Payment Mandate's \texttt{transaction\_id} and the Closed Checkout Mandate's \texttt{checkout\_hash} are both hashes of \texttt{checkout\_jwt}. This common digest binds both mandates to the merchant-signed checkout, so applying a Payment Mandate to another checkout fails verification. In HNP flows, a mandate chain also binds each closed mandate to its parent open mandate.

\subsubsection{Receipts}

The Merchant returns a Checkout Receipt, and the payment-processing side returns a Payment Receipt. Each records acceptance or an error and is bound to its closed mandate. The mandates and receipts form the transaction evidence trail.

\subsubsection{HP Flow}

In HP mode, the user is present throughout the transaction and
directly approves both closed mandates. The flow proceeds as follows~\cite{ap2spec}.
\begin{enumerate}
    \item The user initiates a session with the SA.
    \item The SA communicates with the Merchant and assembles a cart.
    \item Upon checkout, the Merchant creates and signs \texttt{checkout\_jwt}. The hash of this signed checkout is included in the closed Checkout and Payment Mandates.
    \item The SA retrieves the available payment options from the CP and selects one.
    \item The SA constructs the Checkout and Payment Mandate content and requests approval via the TS.
    \item The TS renders the content to the user and obtains consent. The resulting mandates are signed under the applicable AP2 authorization model, using either a User Credential or a key held by a trusted Agent Provider. The \texttt{checkout\_hash} in the Checkout Mandate and the \texttt{transaction\_id} in the Payment Mandate carry the same checkout digest.
    \item The SA passes the signed Payment Mandate to the CP, which verifies it and issues a payment token.
    \item The SA presents the Checkout Mandate and token to the Merchant, which verifies the checkout against what it originally signed and initiates payment via the MPP.
    \item On completion, a Checkout Receipt is provided to the SA, and a Payment Receipt is provided to the SA, CP, and any involved network.
\end{enumerate}

\subsubsection{HNP Flow}

HNP mode differs from the HP flow in two ways. First, the user authorizes the SA through open mandates before leaving the session. Second, the SA signs the corresponding closed mandates autonomously with its own key rather than routing them through the TS. The flow proceeds as follows~\cite{ap2spec}.
\begin{enumerate}
    \item The SA assembles open Checkout and Payment Mandate content for the shopping session and requests approval via the TS.
    \item The TS renders the open mandate content and obtains the user's authorization. The resulting open mandates are signed under the applicable AP2 authorization model. Each open mandate includes the agent's public key (\texttt{agent\_pk}) as a confirmation claim, constraining the mandate to that SA.
    \item The user leaves the session. 
    \item The SA communicates with the Merchant and assembles a cart. The Merchant signs \texttt{checkout\_jwt} as in HP.
    \item The SA selects the open mandates whose constraints cover the assembled checkout, constructs the closed Checkout and Payment Mandates, and signs them with \texttt{agent\_sk}. The \texttt{checkout\_hash} and \texttt{transaction\_id} claims carry the same checkout digest, while the mandate chain binds each closed mandate to its parent open mandate.
    \item The SA presents the open and closed Payment Mandates to the CP and the corresponding Checkout Mandates to the Merchant. Each verifier validates the relevant signatures, bindings, and open mandate constraints. Credential issuance, settlement, and receipt generation then proceed as in HP.
\end{enumerate}

\subsubsection{Phase Segmentation}
\label{subsec:phases}

The AP2~v0.2 distinctions between HP (Direct) and HNP (Autonomous) modes and between open and closed mandates are too coarse for threat modeling~\cite{ap2spec}. We therefore divide a transaction into five phases based on how mandates are prepared, formed, authorized, consumed, and retained. P1 establishes the identities, keys, endpoints, credentials, and trust anchors needed for mandate operations. P2 assembles the cart, payment options, and merchant-signed checkout. P3 converts intent and constraints into signed mandates. P4 verifies and consumes them to authorize payment and settlement. P5 retains mandates, checkout records, and receipts for audits and disputes.
These phases are analytical stages, not a chronology shared by both modes. P3-U denotes user authorization through the TS. P3-A is the HNP-only stage in which the SA derives and signs closed mandates constrained by the user's open mandates.
Figure~\ref{fig:phase-order} shows the order in each flow. HP is sequential; in HNP, open-mandate signing (P3-U) precedes cart assembly (P2). Table~\ref{tab:phase-decomposition} maps each phase to both flows.

\usetikzlibrary{arrows.meta}
\begin{figure}[t]
\centering
\resizebox{\columnwidth}{!}{%
\begin{tikzpicture}[
  phase/.style={rounded corners=3pt, line width=0.7pt, minimum width=1.1cm,
                minimum height=0.7cm, align=center, inner sep=1pt, font=\small\bfseries},
  p1/.style={phase, fill=blue!12,   draw=blue!60},
  p2/.style={phase, fill=orange!22, draw=orange!75},
  p3/.style={phase, fill=green!18,  draw=green!60},
  p4/.style={phase, fill=red!13,    draw=red!60},
  p5/.style={phase, fill=black!8,   draw=black!50},
  pname/.style={font=\scriptsize, text=black!70, align=center, text width=1.4cm, anchor=north},
  lane/.style={font=\bfseries, anchor=east, text=black!80},
  flow/.style={-{Stealth[length=3mm,width=2.6mm]}, line width=0.9pt, draw=black!75}
]

\node[lane] at (-1.0,0) {HP};
\node[p1] (hp1) at (0,0)    {P1};
\node[p2] (hp2) at (1.55,0) {P2};
\node[p3] (hp3) at (3.1,0)  {P3-U};
\node[p4] (hp4) at (4.65,0) {P4};
\node[p5] (hp5) at (6.2,0)  {P5};
\foreach \a/\b in {hp1/hp2,hp2/hp3,hp3/hp4,hp4/hp5} \draw[flow] (\a.east)--(\b.west);
\node[pname] at (0,-0.46)    {Preparation};
\node[pname] at (1.55,-0.46) {Context\\formation};
\node[pname] at (3.1,-0.46)  {User\\signing};
\node[pname] at (4.65,-0.46) {Verification};
\node[pname] at (6.2,-0.46)  {Retention};

\node[lane] at (-1.0,-1.9) {HNP};
\node[p1] (hn1) at (0,-1.9)    {P1};
\node[p3] (hnu) at (1.55,-1.9) {P3-U};
\node[p2] (hn2) at (3.1,-1.9)  {P2};
\node[p3] (hna) at (4.65,-1.9) {P3-A};
\node[p4] (hn4) at (6.2,-1.9)  {P4};
\node[p5] (hn5) at (7.75,-1.9) {P5};
\foreach \a/\b in {hn1/hnu,hnu/hn2,hn2/hna,hna/hn4,hn4/hn5} \draw[flow] (\a.east)--(\b.west);
\node[pname] at (0,-2.36)    {Preparation};
\node[pname] at (1.55,-2.36) {User\\signing};
\node[pname] at (3.1,-2.36)  {Context\\formation};
\node[pname] at (4.65,-2.36) {Agent\\signing};
\node[pname] at (6.2,-2.36)  {Verification};
\node[pname] at (7.75,-2.36) {Retention};
\end{tikzpicture}%
}
\caption{Phase ordering in the two AP2 flows (Table~\ref{tab:phase-decomposition}). Phases are analytical stages, not a fixed chronology.}
\label{fig:phase-order}
\end{figure}
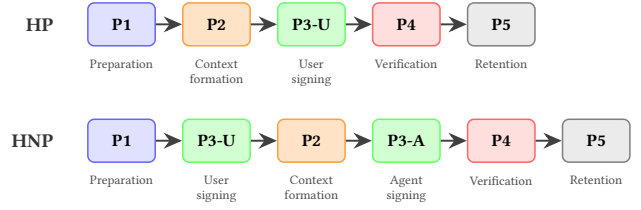

\begin{table*}[!t]
\centering
\caption{Five analytical phases of the AP2 transaction lifecycle.}
\label{tab:phase-decomposition}
\footnotesize
\setlength{\tabcolsep}{2.5pt}
\renewcommand{\arraystretch}{1.00}
\begin{tabularx}{\textwidth}{
  @{}
  >{\raggedright\arraybackslash}p{0.25\textwidth}
  >{\raggedright\arraybackslash}p{0.60\textwidth}
  >{\raggedright\arraybackslash}X
  @{}
}
\toprule
\rowcolor{black!10}
\textbf{Phase} &
\textbf{Lifecycle function} &
\textbf{Flow mapping} \\
\midrule

\textbf{P1} Pre-Mandate Preparation &
Before transaction-specific mandates exist, identities, signing keys,
credentials, AgentCards (A2A discovery documents listing agent identity,
endpoints, and capabilities), endpoint bindings, and trust anchors are
established for mandate creation and verification. &
Precondition for HP and HNP. \\

\rowcolor{black!4}
\textbf{P2} Mandate Context Formation &
The cart, payment options, and merchant-signed checkout are assembled
for closed mandate formation. In HNP, prior open mandates constrain this
context. &
HP steps 1--4; HNP step 4. \\

\textbf{P3} Mandate Authorization &
Unsigned intent or constraints $\rightarrow$ signed mandates. P3-U creates
user-signed closed mandates in HP and user-signed open mandates in HNP;
P3-A derives SA-signed HNP closed mandates constrained by their parents. &
P3-U: HP steps 5--6 and HNP steps 1--2. HNP step 3 transitions from
P3-U to P2. P3-A: HNP step 5. \\

\rowcolor{black!4}
\textbf{P4} Mandate Verification \& Consumption &
Signed, unconsumed mandates $\rightarrow$ verified signatures, mandate
chain, constraints, checkout binding, expiration, and prior-consumption
status, followed by credential issuance, settlement, and receipts. &
HP steps 7--9; HNP step 6. \\

\textbf{P5} Post-Transaction Retention &
After execution, completed mandates, checkout records, and receipts are
integrity-protected and retained as audit and dispute evidence. &
Postcondition for both flows, beginning after receipt generation. \\

\bottomrule
\end{tabularx}
\end{table*}

\subsubsection{Security Requirements and Baseline Assumptions}
\label{subsec:security-requirements}
Our threat model separates three baseline assumptions (BA1--BA3), which are outside our scope, from eight security requirements (SR1--SR8), whose violations we evaluate. For the TS, BA3 assumes only that it is not LLM-driven, whereas SR1 requires faithful mandate rendering and authorization capture.

\paragraph{Baseline Assumptions}
\begin{description}
    \item[{BA1.}] Standard cryptographic primitives are sound when used: signature schemes are unforgeable and hash functions are collision resistant. Parameter adequacy is a deployment property and remains in scope under T-38.
    \item[{BA2.}] Implementations follow all AP2-defined requirements for cryptographic constructions and message formats; attackers therefore cannot forge signatures, substitute mandates, or bypass required bindings. This assumption covers only AP2-defined properties. We evaluate implementer-controlled properties, including salt generation, key lifetime, and optional-field integrity.
    \item[{BA3.}] The TS is not LLM-driven.
\end{description}

\paragraph{Security Requirements}
\begin{description}
    \item[{SR1.}] \textbf{Faithful TS rendering and authorization capture.} The TS faithfully renders mandate content to the user and captures the user's authorization action.
    \item[{SR2.}] \textbf{Signer attribution.} Each mandate is attributable to its signing key, and the verifier establishes that the key is authorized for the relevant role. A mandate created through the TS represents user authorization only if the verifier validates the applicable User Credential or trusts the Agent Provider responsible for the TS. In HNP, a closed mandate signed with \texttt{agent\_sk} represents user authorization only if it is bound to the agent key named in a valid, user-authorized open mandate and satisfies that mandate's constraints.
    \item[{SR3.}] \textbf{Independent verifier checks.} Each verifier independently performs the signature, binding, and constraint checks assigned to its protocol role rather than relying solely on another party's assertion.
    \item[{SR4.}] \textbf{Checkout and mandate-chain binding.} The Closed Checkout Mandate carries the digest of \texttt{checkout\_jwt} in \texttt{checkout\_hash}; the Closed Payment Mandate carries the same digest in \texttt{transaction\_id}. Verifiers confirm that both values bind the mandates to the same merchant-signed checkout, compute every required hash over the AP2-prescribed representation, and reject mismatches. In HNP, each closed mandate also commits to its parent open mandate.
    \item[{SR5.}] \textbf{HNP constraint enforcement.} In HNP, verifiers enforce each open mandate's constraints when deciding whether a closed mandate signed with \texttt{agent\_sk} remains within the user-approved limits. This requirement does not assume that the encoded constraints faithfully represent the user's intent.
    \item[{SR6.}] \textbf{Key confidentiality and configuration integrity.} Signing keys remain confidential to authorized holders and bound to their permitted roles. Credentials, AgentCards, registry records, endpoint configurations, and other security information established during preparation remain authentic.
    \item[{SR7.}] \textbf{Replay prevention and single-use authorization.} Verifiers reject expired or replayed closed mandates, each of which may authorize execution only once. An open mandate may authorize multiple closed mandates only when its recurrence and budget constraints permit reuse and all cumulative limits remain satisfied.
    \item[{SR8.}] \textbf{Evidence retention and later verification.} Signed mandates, checkout records, and receipts needed for disputes remain available, attributable, and protected against tampering.
\end{description}

\textit{Relationship to the Lifecycle Phases.}
The lifecycle maps each requirement to changes in mandate state: key and credential preparation in P1, mandate formation and authorization in P2--P3, verification and consumption in P4, and evidence retention in P5. Failures of key binding, constraint enforcement, serialization, or hash computation are attributed to the phase containing the affected mandate operation and remain in scope.

\subsection{AP2 Deployment Architectures}
\label{subsec:ap2-deployment-arch}

In the AP2-based deployments studied here, A2A carries shopping and checkout messages between agents, while MCP may expose their tools and external services~\cite{ap2spec}. Agentic and deterministic roles, A2A patterns, and MCP configurations yield many architectural variants. We focus on agentic roles to characterize threats from LLM-mediated behavior. We define a distinct architecture class when a deployment introduces a new cross-role trust boundary, a new shared state domain, a new tool-execution surface, or a compromise path that reaches a different set of AP2 roles. Each such class exposes at least one threat that no already-defined class exposes. A configuration that only combines the surfaces of existing classes, and therefore exposes no threat beyond their union, is treated as a composition rather than a new class. Applying this rule yields five architecture classes.

\begin{description}
\item[{A1: Single-Agent SA and Single-Agent MA.}] A1 is our baseline and matches Google's official HP samples~\cite{ap2spec}. One LLM-driven SA communicates with one LLM-driven MA over A2A; the CP and MPP are independent peers reachable through A2A or direct API calls. A1 has no MCP server, internal sub-agent delegation, or marketplace layer. Adversarial input reaches the SA through merchant-side A2A, user input, CP/MPP responses, and external discovery content. Merchant-side A2A is the primary transaction-formation channel because the SA uses it to assemble the cart and negotiate the checkout used in mandate construction. The other channels expand the attack surface without adding a trust boundary.

\item[{A2: Single-Agent SA and Single-Agent MA with Isolated MCP.}] A2 retains A1's single-agent structure but adds an isolated MCP server to each side. The servers share no state, cache, storage, credentials, or authorization context across the SA/MA boundary. The added tool-execution surface exposes MCP threats absent from A1 even though it adds no SA/MA boundary.

\item[{A3: Multi-Agent SA and Multi-Agent MA.}] A3 implements the SA and MA as specialized sub-agents. An SA may have monitoring, purchase, and consent sub-agents under an orchestrator, while an MA may have catalog, pricing, and checkout sub-agents. The external AP2 interface remains unchanged, with the SA communicating with the MA over A2A. A3 introduces vulnerabilities from malicious sub-agent propagation and orchestration failures. Adding isolated MCP yields A3$\times$A2, a composition rather than a separate class because it exposes no threat beyond the union of A3 and A2.

\begin{figure}[t]
    \centering    \includegraphics[width=0.98\columnwidth]{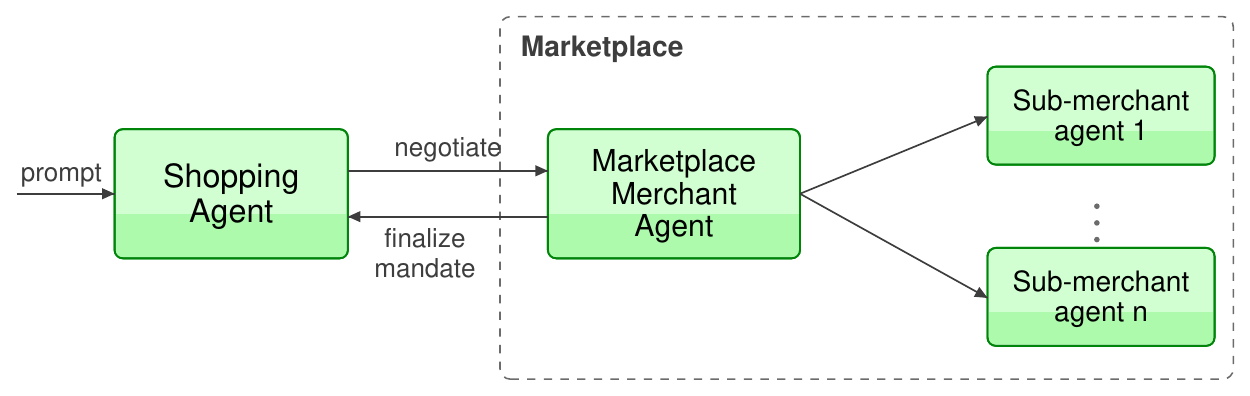}
    \caption{A4: Marketplace architecture.}
    \label{fig:marketplace}
\end{figure}

\item[{A4: Marketplace.}] In A4, shown in Figure~\ref{fig:marketplace}, the SA interacts with one marketplace identity that routes transactions to sub-merchants and mediates the AP2 lifecycle. During discovery and cart assembly, the SA uses one marketplace-facing agent instead of negotiating with each merchant as an independent AP2 counterparty. Behind that identity, the marketplace resolves the user's request internally by selecting one or more sub-merchants, catalog entries, fulfillment options, and fee rules.

Unlike A1--A3, A4 can separate the visible AP2 merchant identity from the sub-merchant supplying the item, price, fulfillment state, or dispute facts. A user may approve a marketplace-scoped mandate while hidden actors handle execution and evidence, creating an opaque merchant-side trust boundary. Using one or many internal agents changes compromise scale but not the class. Shared MCP produces an A4$\times$A5 composition because it combines the marketplace and shared-MCP boundaries without exposing threats beyond their union.

\begin{figure}[t]
    \centering    \includegraphics[width=0.98\columnwidth]{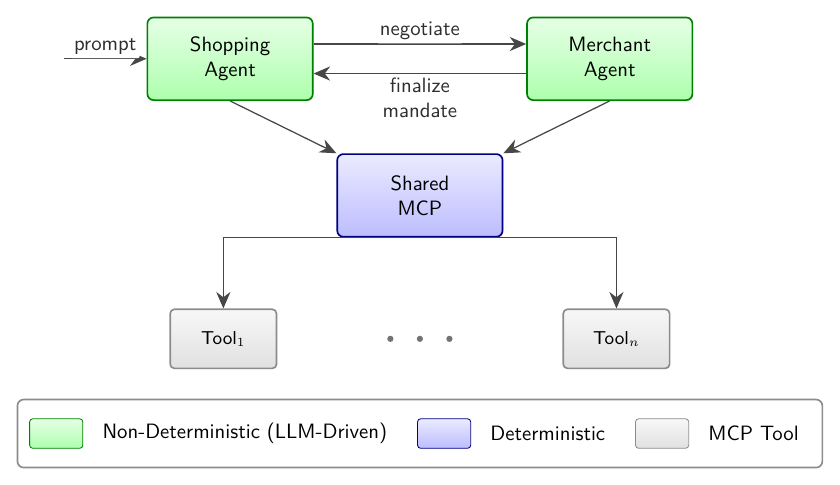}
    \caption{A5: Shared MCP architecture.}
    \label{fig:shared}
\end{figure}

\item[{A5: Shared MCP.}] A5, shown in Figure~\ref{fig:shared}, 
retains the SA--MA A2A channel but gives both roles the same MCP server during discovery, cart assembly, and checkout preparation. Unlike architectures with no tools or isolated tool use, this server is an operational surface and influence point for both roles. The SA may search products, compare offers, or construct a candidate cart, while the MA may resolve inventory, pricing, merchant policies, cart state, or checkout records. AP2 messages still use A2A, but some message state and resulting artifacts derive from a shared tool layer that may produce product data, price quotes, cart identifiers, and tool results. A compromised or malicious server can therefore influence both transaction sides through common MCP attacks. Internal agent variations change the compromise footprint but not the class.

\end{description}

\vspace{-1em}
\section{Threat Model}
\label{sec:threat-model}

We characterize each threat as a tuple of four elements: the actor initiating it, the surface on which they act, the capability they wield, and their goal. Phase applicability is recorded per threat in Figure~\ref{fig:threat-matrix} rather than in the tuple, since several threats span phases (e.g., P2--P4).

Threats are identified using MAESTRO~\cite{maestro2025}, whose seven layers cover foundation models (L1), data operations (L2), agent frameworks (L3), deployment infrastructure (L4), evaluation and observability (L5), security and compliance (L6), and the agent ecosystem (L7). The MAESTRO layers are tagged per surface in Section~\ref{subsec:attack-surfaces} and per threat in Figure~\ref{fig:threat-matrix}. We assign L1 only when a threat depends on foundation-model behavior. We classify a threat that changes MCP or tool metadata as L7, L3, or L2, according to the affected object. Therefore, we assign T-2 to L7 because it changes MCP tool metadata. We assign T-1 to L1 because it exploits the model's response to poisoned context.

\textit{Notation}
Access capabilities use the prefix AC followed by a family letter. We use K for key access, M for mandate and token access, C for communication-channel access and R for runtime access. Knowledge capabilities use the prefix AK. Surfaces are labeled S1--S11, threat actors are labeled TA1--TA4, and attacker goals are labeled AG1--AG6. We refer to the five deployment architectures described in Section~\ref{subsec:ap2-deployment-arch} as A1-A5 and to the five protocol phases described in \ref{subsec:phases} as P1--P5. Security requirements are labeled SR1--SR8 and baseline assumptions BA1--BA3; both are defined in Section~\ref{subsec:security-requirements} and cited by number throughout.

\subsection{Attacker Model}
\label{subsec:attacker-model}

An attacker is characterized by their access capabilities, which describe what they can read, write, invoke or control, and knowledge capabilities, which describe what they know about a deployment. Access capabilities are required to realize a threat. Knowledge capabilities do not realize threats by themselves, but they reduce attack complexity or increase the probability of success. The capability codes used in the threat model are summarized in Table~\ref{tab:actor-capability-summary}.

A compromised component gives the attacker the same permissions and access that component has. If several components are compromised, their permissions combine. In Architecture A5, both the SA and MA call the same MCP server, so compromising that server may let an attacker interfere with both agents’ tool interactions in a single transaction. Therefore, threats must be analyzed by system architecture, not just by agent role.

\begin{table*}[t]
\centering
\footnotesize
\setlength{\tabcolsep}{4pt}
\renewcommand{\arraystretch}{1.00}
\caption{Summary of the capability codes used in the AP2 threat model. 
\texttt{I} means inherent to the actor in normal operation, 
\texttt{C} means obtainable through compromise or reachability, and 
\texttt{-} means not typically held.}
\label{tab:actor-capability-summary}

\resizebox{\textwidth}{!}{%
\begin{tabularx}{\textwidth}{
    @{}
    >{\ttfamily}p{0.075\textwidth}
    >{\raggedright\arraybackslash}X
    c c c c
    @{}
}
\toprule
\normalfont\textbf{Code} & \textbf{Meaning} &
\textbf{TA1} & \textbf{TA2} & \textbf{TA3} & \textbf{TA4} \\
\midrule

\multicolumn{6}{@{}l}{\textbf{Keys}} \\
ACK1 & Control of \texttt{user\_sk}; can produce user-signed mandates.
     & I & - & C & C \\
ACK2 & Control of \texttt{agent\_sk}; can produce HNP closed mandates.
     & C & - & C & C \\
ACK3 & Control of \texttt{merchant\_sk}; can sign checkout artifacts and receipts.
     & - & I & C & C \\
ACK4 & Control of CP/MPP verifier keys; can forge payment tokens or receipts.
     & - & - & I & C \\

\midrule
\multicolumn{6}{@{}l}{\textbf{Mandates \& Tokens}} \\
ACM1 & Access to public mandate, token, or receipt claims.
     & I & I & I & C \\
ACM2 & Access to selectively disclosed mandate or receipt claims.
     & I & I & I & C \\
ACM3 & Access to a scoped payment token in flight.
     & - & C & C & C \\

\midrule
\multicolumn{6}{@{}l}{\textbf{Channels}} \\
ACC1 & Passive observation of A2A or MCP traffic.
     & - & C & C & I$^{b}$ \\
ACC2 & Active modification of A2A or MCP messages in transit.
     & - & C & C & C \\

\midrule
\multicolumn{6}{@{}l}{\textbf{Runtime}} \\
ACR1 & RPC-level access to an exposed AP2 role or tool interface.
     & I & I & I & C \\
ACR2 & Process-owner access to memory, local state, or verifier logic.
     & I$^{a}$ & I & I & C \\
ACR3 & Host-level control of the target role or infrastructure.
     & I$^{a}$ & I & I & C \\
ACR4 & Build-pipeline or release influence on AP2 software.
     & - & I & I & C \\
ACR5 & MCP-server-operator access; control over tool descriptions or tool results.$^{c}$
     & - & C & I & C \\

\midrule
\multicolumn{6}{@{}l}{\textbf{Knowledge}} \\
AK1 & Knowledge of deployed architecture, agentic roles, and shared infrastructure.
    & C & C & I & C \\
AK2 & Knowledge of exposed tools, capabilities, and MCP resources.
    & - & I & I & C \\
AK3 & Knowledge of the LLM model, version, or agent framework.
    & - & I & I & C \\
AK4 & Knowledge of mandate contents, constraints, or disclosed claims.
    & I & I & I & C \\

\bottomrule
\multicolumn{6}{@{}p{\textwidth}@{}}{\footnotesize
$^{a}$~TA1 inherency for ACR2/ACR3 assumes a user-controlled SA host; under
hosted-agent deployments these become C and the corresponding runtime
capability shifts to TA3.
$^{b}$~TA4's inherent ACC1 reflects an assumed on-path network vantage;
absent that vantage it is C.
$^{c}$~ACR5's blast radius is architecture-conditional: in A2 and
A3$\times$A2 it influences one role's tool calls; in A5 and A4$\times$A5
(shared server) it influences both.
}
\end{tabularx}%
}
\end{table*}

\subsection{Threat Actors}
\label{subsec:threat-actors}

A threat actor 
may be an individual, a group, or an organization. Each actor is assigned a security posture: honest, honest-but-curious, or malicious for TA1--TA3, and malicious by assumption for TA4. Honest postures are included so the actor set also identifies principals appearing as victims, counterparties, or compromised surfaces in a given threat. We derive four threat actors by grouping the AP2 roles defined in Section~\ref{subsec:roles}, along with the User principal, according to their adversarial objectives.

\begin{description}
\item[{[TA1] User.}]
The customer who initiates the transaction and authorizes consent at the TS. An honest-but-curious User probes to infer verifier policy or merchant-side state that was not intentionally exposed. A malicious User on the other hand repudiates consent, or tampers the credential store, or submits transaction inputs they later dispute. The User has AK4 for their own mandates and inherently holds ACK1, along with ACR2--ACR3 on a user-controlled SA host, but does not have build-pipeline access (ACR4).

\item[{[TA2] Merchant.}]
The legal entity occupying the Merchant role, Merchant Agent, Merchant BE, or merchant MCP server. A malicious Merchant may overcharge, substitute items, manipulate catalog content, or return tool output that steers mandate construction away from the user’s intent. The Merchant inherently holds ACK3, ACR2--ACR4 over its own infrastructure and release builds. In A4, the Merchant role may comprise multiple merchant-side principals, including a marketplace operator and one or more sub-merchants. We model these as instances of the same actor class, TA2, rather than as separate actor labels. When the distinction matters, we identify the responsible principal from the affected S4 backend: a marketplace-front backend points to the marketplace operator, a sub-merchant backend to a sub-merchant, and a shared backend to shared merchant infrastructure.

\item[{[TA3] Developer / Operator.}]
Parties that implement, deploy, or operate AP2 surfaces, including SA providers,
TS vendors, CP and MPP operators, and registry operators. A malicious operator may ship a backdoored release, weaken verifier logic, exfiltrate keys, or route traffic through an unintended dependency. Per operated surface, this actor holds the relevant key, runtime, release and configuration capabilities. 

\item[{[TA4] External Attacker.}]
Any principal outside the named transaction parties and operators, including a network attacker, supply-chain attacker, on-host malware operator, or third-party MCP provider outside the deployment trust boundary. Its capabilities come from reachable or compromised surfaces, not from a default AP2 role. 

\item[{Modeling choices.}]
The model assigns intent to threat actors, not to surfaces. We therefore model the TS, SA, MA, CP, MPP, and AgentCard registry as surfaces (S1--S7), not as threat actors. We attribute malicious behavior to the actor that operates or compromises the surface (TA1--TA4, as applicable). When actors collude, we combine their capabilities and assign their shared goal. We do not define a separate actor class for collusion.
\end{description}

\subsection{Attack Surfaces}
\label{subsec:attack-surfaces}

An attack surface is a point at which a threat actor's capabilities can affect AP2 security. We identify 11 AP2 attack surfaces and organize them into three categories: role and runtime surfaces (S1--S7), communication-channel surfaces (S8--S9), and data and artifact surfaces (S10--S11). These surfaces are described in Table~\ref{tab:attack_surfaces_exhaustive}.

One surface, the S9 MCP channel, varies across architectures: it is absent in A1, pure A3, and pure A4; isolated per side in A2 and A3×A2; and shared in A5 and A4×A5. We nonetheless keep it as a single surface, so the deployment architecture determines whether an S9 threat applies and whether its scope is confined to a single side or spans both. The tuple's architecture element records the applicable case.

\begin{table*}[!]
\caption{Systemic AP2 attack surfaces ($S_1$--$S_{11}$).}
\label{tab:attack_surfaces_exhaustive}
\centering
\footnotesize
\setlength{\tabcolsep}{3.5pt}
\renewcommand{\arraystretch}{1.06}
\begin{tabularx}{\textwidth}{@{}>{\raggedright\arraybackslash}p{0.13\textwidth} >{\raggedright\arraybackslash}X >{\raggedright\arraybackslash}X >{\raggedright\arraybackslash}p{0.13\textwidth}@{}}
\toprule
\textbf{Surface} & \textbf{Core role} & \textbf{Risk / failure mode} & \textbf{MAESTRO} \\ 
\midrule

\multicolumn{4}{l}{\textbf{Role and Runtime Surfaces}} \\
\midrule

\textbf{S1: TS} & 
UI surface trusted to render mandate content and obtain user consent before signing. & 
Mandate mismatch, consent bypass, or local key exposure (\texttt{user\_sk}, \texttt{ACR2--4}). & 
L4, L6 \\
\specialrule{0.1pt}{0.4pt}{0.4pt}

\textbf{S2: SA} & 
SA for cart assembly, mandate construction, and payment execution. & 
Prompt injection, jailbreaks, or tool poisoning altering mandate generation fields. & 
L1--L4, L6, L7 \\
\specialrule{0.1pt}{0.4pt}{0.4pt}

\textbf{S3: MA} & 
LLM-driven MA interacting with the SA over A2A. & 
Catalog/quote manipulation or tool tampering; signing compromise maps to \texttt{ACK3}. & 
L1--L4, L6, L7 \\
\specialrule{0.1pt}{0.4pt}{0.4pt}

\textbf{S4: Merchant BE} & 
Deterministic merchant backend system. & 
\texttt{merchant\_sk} exposure, checkout state alteration, or marketplace isolation failure (A4: marketplace-front vs.\ sub-merchant backend instance). &
L4, L6 \\
\specialrule{0.1pt}{0.4pt}{0.4pt}

\textbf{S5: CP} & 
Scoped credential release authority. & 
Mandate check bypass, token leakage, or agentic prompt injection. & 
L4, L6 (Agentic: L1--L3, L7) \\
\specialrule{0.1pt}{0.4pt}{0.4pt}

\textbf{S6: MPP} & 
Settlement verification and receipt signer. & 
Settlement/token corruption, faulty receipt generation, or audit tampering. & 
L4, L6 (Agentic: L1--L3, L7) \\
\specialrule{0.1pt}{0.4pt}{0.4pt}

\textbf{S7: Registry} & 
Identity and discovery infrastructure (AgentCard). & 
Impostor routing, extension suppression, or discovery provenance manipulation. & 
L4, L7 \\ 
\midrule

\multicolumn{4}{l}{\textbf{Communication-Channel Surfaces}} \\
\midrule

\textbf{S8: A2A Channel} & 
Inter-agent messaging substrate. & 
Protocol Downgrade, endpoint hijacking, or conversational state replays. & 
L4, L6, L7 \\
\specialrule{0.1pt}{0.4pt}{0.4pt}

\textbf{S9: MCP Channel} & 
Tool execution interface. & 
Tool poisoning, result tampering, or cross-role state bleeding (shared server A5). &
L2--L4, L6, L7 \\
\midrule

\multicolumn{4}{l}{\textbf{Data and Artifact Surfaces}} \\
\midrule

\textbf{S10: Artifacts} & 
Signed credentials and evidence (SD-JWTs, tokens, receipts, audits). & 
Information leakage, disclosure confusion, stale replay, or token capture. & 
L2, L5, L6 \\
\specialrule{0.1pt}{0.4pt}{0.4pt}

\textbf{S11: Knowledge} & 
External data, catalog caches, and retrieval databases. & 
Ingestion/retrieval poisoning, cross-tenant bleeding, or corpus modification. & 
L1, L2, L6 \\ 
\bottomrule
\end{tabularx}
\end{table*}

\subsection{Attacker Goals}
\label{subsec:attacker-goals}

We summarize the attacker objectives as six goals. Each goal is realized by one or more threats. Targets denotes the security properties whose violation can realize the goal; 
SR3 (independent verification) is a latent target for any goal requiring a malformed or out-of-scope artifact to pass verification; we list it explicitly only where it is the primary failure.

\begin{description}
\item[{[AG1] Unauthorized Payment.}] Cause settlement with no valid corresponding authorization (forged, absent, or invalid), such as a payment the user never approved or approved for a different amount, merchant, or instrument. Targets SR2, SR4, and SR6.

\item[{[AG2] Mandate-Content Manipulation.}] Obtain a valid signature over content whose effective semantics differ from the user's intent, rendered view, or pre-authorized constraints, such as a hidden charge, substituted item, or relaxed allowed\_merchants list; the manipulation occurs before or during authorization, including mandate construction, rendering, constraint encoding, or signing mediation. Targets SR1, SR4, and SR5.

\item[{[AG3] Authorization-Scope Inflation.}] Extend a legitimately constructed authorization beyond its scope after signing (replayed, reused, or rebound). Targets SR3, SR4, SR5 and SR7.

\item[{[AG4] Confidentiality Breach.}] Recover selectively disclosed claims, observe payment metadata, or link transactions across the dispute-evidence retention window. Targets the confidentiality of S10 and communication-channel metadata on S8/S9.

\item[{[AG5] Repudiation.}]
Deny a legitimately authorized transaction through weaknesses in evidence retention or mandate-chain auditability. Targets SR2 and evidence retention in P5.

\item[{[AG6] Denial of Authorized Payment.}] Prevent an authorized transaction from being completed, for example, by misrouting discovery, or stalling MCP tool calls.  
Targets availability of the authorized transaction across P2--P4 (SR3/SR4 when caused by verifier or binding failure).
\end{description}

\subsection{Risk Assessment}
\label{subsec:threat-tuple}
Risk is assessed separately for each deployment architecture. Each threat uses the tuple $\langle\mathrm{actor},\allowbreak\,\mathrm{surface},\allowbreak\,\mathrm{capability},\allowbreak\,\mathrm{architecture},\allowbreak\,\mathrm{goal}\rangle$ defined above and applies only when the architecture exposes the required surface and its preconditions hold. We score inherent risk using AIVSS v0.5~\cite{aivss}, which combines a CVSS v4.0 base score~\cite{cvss40} with Agentic AI Risk Score (AARS) amplification factors. We assume PoC threat maturity and apply no mitigation discount, ranking inherent risk before deployment-level controls. A threat is high if its AIVSS score reaches the high-severity band in at least one architecture. The complete per-threat breakdown for all 48 threats along with their full AIVSS score table and the full threat matrix is provided in the companion repository.\footnote{\url{https://anonymous.4open.science/r/AP2_Beyond_the_Mandate}}

\subsubsection{Scoring Method}
\label{subsec:aivss-risk-assessment}

We score each applicable threat in two layers. First, we assign a CVSS~v4.0 base score~\cite{cvss40} from the protocol-level attacker path. Exploitability is derived from the minimum required access capability, the number of prerequisite compromises, and whether the attack requires a timing window, privileged role, or ordinary AP2 transaction flow. Impact is derived from the AP2 attacker goal mapped to CVSS dimensions as shown in Table~\ref{tab:aivss-goal-impact}.

\begin{table}[t]
\centering
\caption{Mapping from AP2 attacker goals to CVSS impact dimensions.}
\label{tab:aivss-goal-impact}
\small
\setlength{\tabcolsep}{3pt}
\renewcommand{\arraystretch}{1.05}
\begin{tabularx}{\linewidth}{@{}lX@{}}
\toprule
\textbf{Goal} & \textbf{CVSS impact interpretation} \\
\midrule
AG1 Unauthorized payment & Integrity impact on the AP2 transaction and subsequent payment system. \\
AG2 Mandate-content manipulation & Integrity impact on mandate semantics and signed user intent. \\
AG3 Authorization-scope inflation & Integrity impact on the scope or reuse of authorization. \\
AG4 Confidentiality breach & Confidentiality impact on disclosed claims, payment metadata, or linkage data. \\
AG5 Repudiation & Integrity impact on evidence, receipts, and dispute artifacts. \\
AG6 Denial of authorized payment & Availability impact on completion of an authorized transaction. \\
\bottomrule
\end{tabularx}
\end{table}

Second, we assign the ten AARS amplification factors on a three-level scale. A value of $0$ means the factor is not applicable to the scored threat path, $0.5$ means it is partially present or deployment-dependent, and $1$ means it is fully present. The factors are autonomy of action, tool use, memory use, dynamic identity, multi-agent interaction, non-determinism, self-modification, goal-driven planning, contextual awareness, and opacity/reflexivity, as defined by AIVSS~\cite{aivss}. AARS is used only as the agentic amplification component inside AIVSS; it is not treated as a separate severity score.

We apply two guardrails when assigning these factors. We reserve non-determinism for attacks in which an LLM decision point contributes to exploit success. We reserve self-modification for cases where the agent can alter its own code, prompt, tool configuration, model configuration, or policy at runtime. Deterministic protocol downgrades, signing-key compromise, registry errors, and historical evidence failures therefore do not receive these factors by default.

The AARS amplification sum is reported on a $0$--$10$ scale:
\[
\mathrm{AARS}=\sum_{i=1}^{10} f_i,\qquad f_i \in \{0,0.5,1\}.
\]

The AIVSS score combines the CVSS~v4.0 base score with the AARS amplification sum, scaled by a threat-maturity multiplier $\mathrm{ThM}$ and a mitigation factor $\mathrm{MF}$:

\[
\mathrm{AIVSS}=\mathrm{round}\!\left(
  \frac{\mathrm{CVSS}_{base}+\mathrm{AARS}}{2}
  \times \mathrm{ThM} \times \mathrm{MF},\;1
\right).
\]

Here $\mathrm{CVSS}_{base}\in[0,10]$ is the protocol-level CVSS~v4.0 base score, $\mathrm{AARS}$ is the factor sum above, $\mathrm{ThM}\in(0,1]$ captures observed threat maturity, and $\mathrm{MF}\in(0,1]$ captures residual exposure after deployed mitigations. The arithmetic mean is used because both CVSS and AARS are on a $0$--$10$ scale; averaging keeps AIVSS bounded within the same range while weighting protocol-level severity and agentic amplification equally. The $\mathrm{round}(\cdot,1)$ operator denotes round-half-up to one decimal place.

Following the AIVSS documentation~\cite{aivss}, we use $\mathrm{ThM}=0.97$ because this assessment evaluates PoCs rather than exploits observed in production deployments. To rank inherent risk before deployment-level controls are applied, we assume the lack of mitigations and set $\mathrm{MF}=1.0$. Substituting these values gives the formula used for every scored row:

\[
\mathrm{AIVSS}=\mathrm{round}\!\left(
  \frac{\mathrm{CVSS}_{base}+\mathrm{AARS}}{2}
  \times 0.97 \times 1.0,\;1
\right).
\]

For example, T-1 (pre-signing context poisoning) has the CVSS vector
\path{CVSS:4.0/AV:N/AC:L/AT:N/PR:N/UI:P/VC:N/VI:H/VA:N/SC:N/SI:H/SA:N},
which gives $\mathrm{CVSS}_{base}=8.3$. Its AARS factor tuple is $(0.5,0.5,0.5,0.5,0.5,1,0,1,1,1)$, so $\mathrm{AARS}=6.5$. Therefore,
\[
\mathrm{AIVSS}=\mathrm{round}\!\left(
  \frac{8.3+6.5}{2}\times 0.97 \times 1.0,\;1
\right)=7.2,
\]
placing the threat in the High band.

\begin{figure*}[!t]
    \centering
    \makebox[\textwidth][c]{%
        \includegraphics[width=1.0\textwidth]{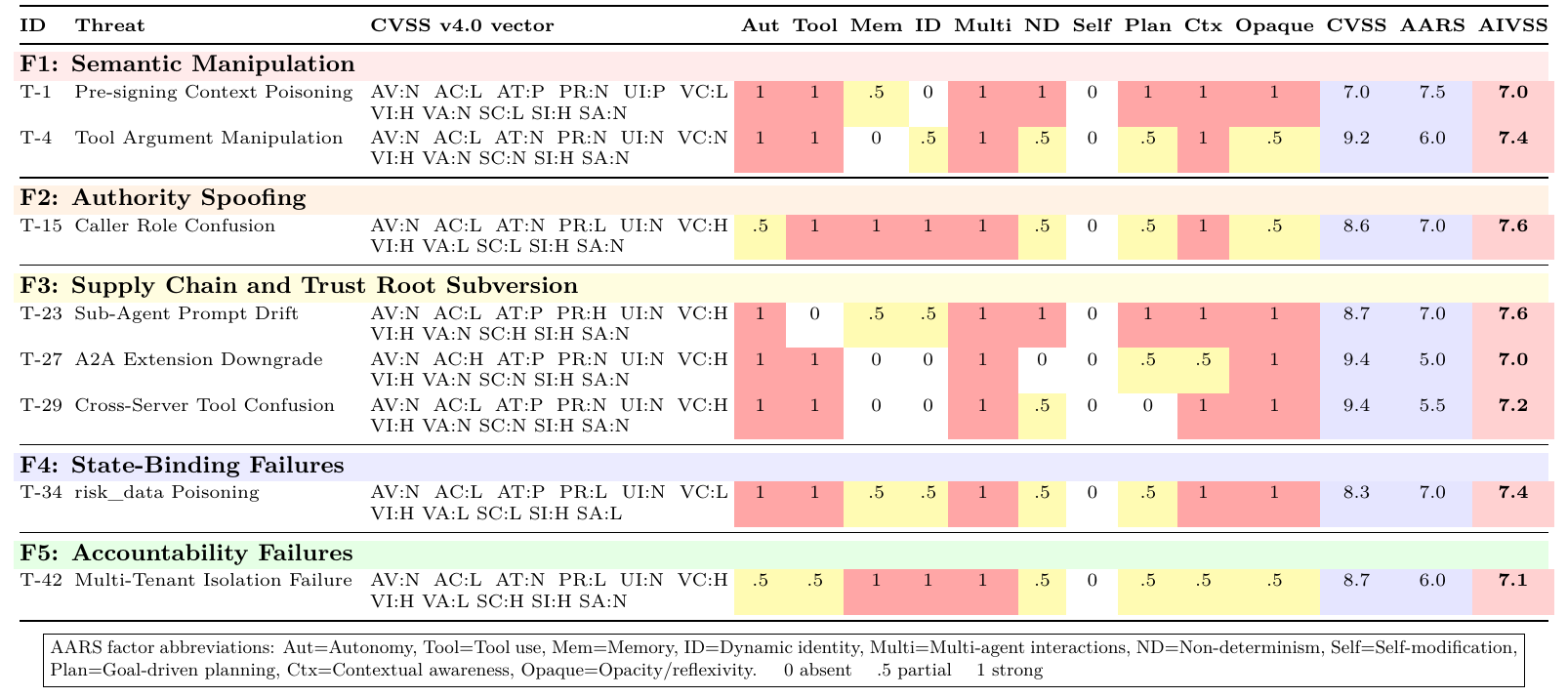}%
    }
    \caption{AIVSS risk scores for the 8 high-risk AP2 threats.}
    \label{fig:aivss-matrix-preview}
\end{figure*}

Rows are labeled using the standard severity bands: None for $0.0$, Low for $0.1$--$3.9$, Medium for $4.0$--$6.9$, High for $7.0$--$8.9$, and Critical for $9.0$--$10.0$. Figure~\ref{fig:aivss-matrix-preview} summarizes the 8 threats that reach the High band, showing each threat's CVSS base score, AARS factor profile, and final AIVSS score.

\subsubsection{Architecture-specific amplification.}
Consider T-31 (mandate replay), which is applicable in both the Single-Agent
architecture (A1) and the Multi-Agent architecture (A3). In A1, one agent can
sequentially reuse a valid mandate. With the factor order used above, this path
has the AARS tuple $(0.5,0.5,1,0,0,0,0,0.5,0.5,1)$, so
$\mathrm{AARS}=4.0$. Holding its $\mathrm{CVSS}_{base}=8.9$ constant gives
$\mathrm{AIVSS}=6.3$. In A3, the same authorization can be fanned out
concurrently across several agents. This activates the multi-agent-interaction
factor while the other factors remain unchanged, producing the tuple
$(0.5,0.5,1,0,1,0,0,0.5,0.5,1)$, $\mathrm{AARS}=5.0$, and
$\mathrm{AIVSS}=6.7$. 

\subsubsection{Threat Matrix reading protocol.}
\begin{figure*}[!t]
  \centering
  \makebox[\textwidth][c]{%
    \includegraphics[width=0.95\textwidth]{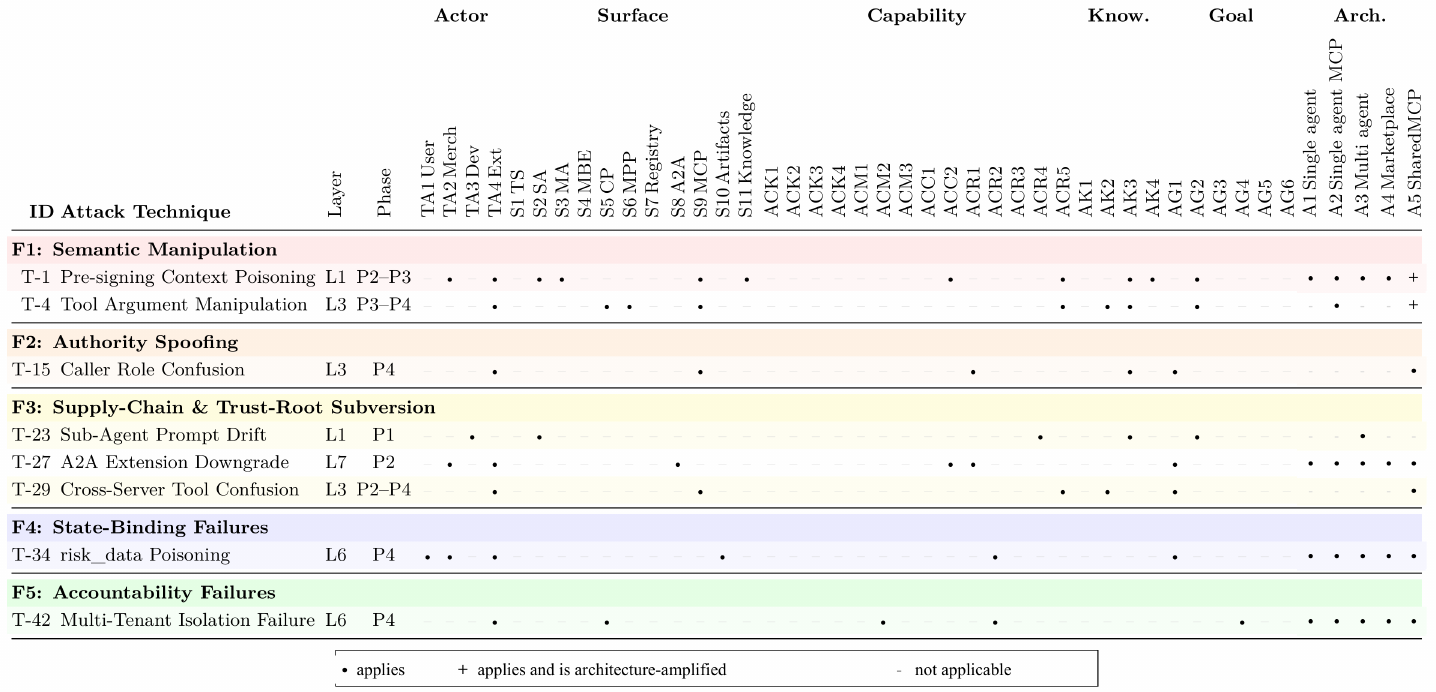}%
  }
  \caption{High risk AP2 threat matrix.}
  \label{fig:threat-matrix}
\end{figure*}

The matrix in Figure~\ref{fig:threat-matrix} can be read in three ways. Reading within an attack-family band shows the common actor, surface, and capability profile for that family. Reading down an architecture column shows which threats become High under that deployment, where the Single-Agent baseline (A1) is dominated by F3 trust-root and F4 state-binding threats, while the Shared-MCP architecture (A5) amplifies the largest number of F1 threats. Reading across a single row reconstructs the threat tuple used to connect the threat model, the risk assessment, and the PoC descriptions.

\section{Taxonomy of AP2 Attack Families}
\label{sec:attack-families}

Using the threat tuple from Section~\ref{subsec:threat-tuple}, we devided the 48 MAESTRO threats into five attack families, each defined by the AP2 \emph{security object} it corrupts. A security object represents the core logical and cryptographic mandates, such as signing authority and transaction state, vital to protocol security. Within each family we further divide the threats into sub-families, where a sub-family shares a common attacker goal and a common fix. Section~\ref{sec:attack-demonstrations} then focuses on the 8 threats that reach the High band by AIVSS. 

\subsection{Classification Methodology}
\label{subsec:family-method}

We categorize the threats using a two-stage bottom-up methodology.
First, we cluster the discovered threats bottom-up based on the specific AP2 security object they corrupt. These objects represent the core logical and cryptographic mandates, such as signing authority and transaction state, vital to protocol security.
Second, we verify these clusters by ensuring that each resulting family shares a distinct AP2 failure mode, overlapping attacker goals, and a common fix. We then refine each family one level down into sub-families, naming each sub-family by its \emph{mechanism} of failure rather than by its symptom.
We rejected alternative taxonomy models because they organize threats by surface-level attributes, such as when the attack occurs, what the attacker ultimately wants, or which capability it uses, rather than by the protocol function that fails:

\begin{description}
    \item \textbf{Phases} merely track an operational timeline rather than exploit mechanics.
    \item \textbf{Goals} are too broad, grouping unrelated vulnerabilities under a single objective.
    \item \textbf{Capabilities} over-fragments the catalog, since identical protocol failures can be reached through entirely different vectors.
\end{description}

These dimensions remain essential for threat attribution and scoring (Sections~\ref{sec:lifecycle} and~\ref{sec:threat-model}); we reject them only as the basis for the family taxonomy, which is organized by corrupted security object.

\begin{figure*}[!t]
    \centering
    \makebox[\textwidth][c]{%
        \includegraphics[width=1\textwidth]{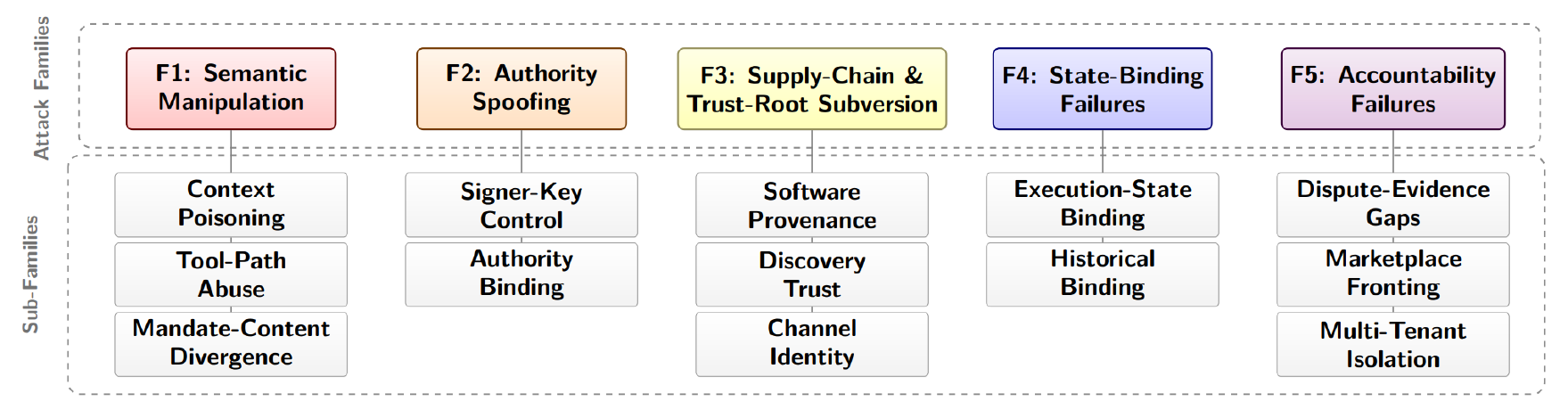}%
    }
    \caption{Taxonomy of AP2 attack families and sub-families.}
    \label{fig:attack_families}
\end{figure*}

This bottom-up process yields five distinct families, each mapped to core security objects: \emph{Semantic Manipulation} (corrupting conversation context), \emph{Authority Spoofing} (corrupting signing authority), \emph{Supply-Chain \& Trust-Root Subversion} (corrupting trust roots), \emph{State-Binding Failures} (corrupting transaction state), and \emph{Accountability Failures} (corrupting accountability records). 
If a threat corrupts more than one security object, we assign it to the object whose corruption constitutes the initial AP2-specific failure, the point at which the protocol's guarantees first break and from which the remaining corruptions follow. For instance, a poisoned tool result under shared MCP (T-1) simultaneously injects adversarial content into the receiving agent's context and exfiltrates transaction state from the originating one. We classify it under Semantic Manipulation rather than State-Binding or Accountability, because the contextual corruption is what first diverts mandate construction from user intent, whereas the state leakage is a downstream confidentiality effect. Figure~\ref{fig:attack_families} summarizes the resulting taxonomy of families and sub-families.

\subsection{F1: Semantic Manipulation}
\label{subsec:f1-semantic-manipulation}

Semantic Manipulation attacks cause AP2 to accept, construct, authorize, or execute mandates whose effective semantics differ from what the user intended to perform. This family contains 10 of the 48 threats (T-1--T-9, T-48). Their primary goal is AG2, with AG1 and AG4 as common consequences. 

\begin{description}
\item[{Context Poisoning.}]
This sub-family targets the agent's working context, the accumulated inputs from which a mandate is formed. By seeding that context with adversarial content before construction, an attacker leads the SA or MA to encode its own intent as though it were the user's. What defines the sub-family is that this single failure mode is reachable through several distinct entry points. The first is pre-signing context poisoning, in which adversarial text arrives via catalog content, an A2A handshake or reply, a RAG corpus, or a cross-zone tool-result fan-out; under shared MCP, that same fan-out both injects content into the receiving zone and leaks state back from the originating one (T-1). The second is MCP metadata poisoning, in which malicious tool descriptions persist and are interpreted with high instruction weight (T-2). The third involves no boundary-crossing injection at all, instead weakening the system prompt through long-context attention dilution or slow multi-turn drift (T-3). The fourth drops injection entirely for persuasion: guardrail-passing language that embeds a malicious intent in the agent's long-running memory or convinces a peer agent to lower its defenses (T-48). Although the vectors may differ, the resulting failure is the same i.e., the generated mandate aligns with the attacker’s context, bypassing the user’s intent.

\item[{Tool-Path Abuse.}]
The attacker does not need to poison the entire conversational history; instead, they can redirect framework’s control flow simply by tampering with tool configurations or schemas. This includes tool argument manipulation, including verifier-side CP-MCP tools that can bias credential release (T-4); tool calls fired before consent or in adversarial order (T-5); and cross-agent semantic collusion, where sub-agent chains compose locally valid actions into a mandate that exceeds user intent (T-9). These attacks exploit the architectural reality that LLM agents treat tool metadata and structural outputs as operational instructions, altering framework outputs or verifier-facing actions to bypass user intent.

\item[{Mandate-Content Divergence.}]
In this sub-family, the mis-match appears inside the cryptographic mandate itself or between the mandate and its user-facing representation, and it can arise at three points in the authorization pipeline. During construction, the SA fills in missing constraints on its own instead of asking the user, producing an over-broad mandate (T-6). At the rendering layer, the user-facing summary diverges from the structured fields that are actually signed (T-7). And inside the SD-JWT disclosure flow, one value is disclosed to the TS while a different signed value is later consumed by the verifier (T-8). Although signature verification succeeds in each case, the protocol still fails, because AP2 binds a mandate the user never authorized.

\end{description}

\subsection{F2: Authority Spoofing}
\label{subsec:f2-authority-spoofing}

Authority Spoofing tricks AP2 verifiers into attributing a mandate, checkout\_jwt or receipt to the wrong signing authority. A legitimate principal appears responsible, although an unauthorized party controlled the signing path or the upstream decision. This family contains 9 of the 48 consolidated threats (T-10--T-17, T-45) and primarily realizes AG1, with specific instances targeting AG2, AG4, and AG5.

\begin{description}
\item[{Signer-Key Control.}]
The attacker obtains control of a long-lived AP2 signing key, or causes that key to be used outside its legitimate authorized scope. This includes signing-key exfiltration of \texttt{user\_sk}, \texttt{agent\_sk}, or \texttt{merchant\_sk} through host or memory compromise (T-10); weak or unsealed key generation, such as low-entropy keygen or an in-memory window before HSM seal (T-11); and MCP credential exposure, where committed OAuth secrets or API keys let an attacker impersonate the operator's MCP client (T-17). In all cases a valid signature proves only that the relevant key was used; it does not establish that the principal it identifies actually approved the action.

\item[{Authority Binding.}]
The attacker exploits an attribution or attestation gap that is missing, too coarse, or checked at the wrong role, caller, or boundary. This includes unanchored agent-key provenance (T-12), unscoped key identifiers that let one role's key verify as another (T-13), sub-agent attestation gaps (T-14), caller-role confusion at shared components (T-15), verifier decisions that leave no proof of checking (T-16), and TS consent-surface hijacking where the user approves a decoy while the real mandate signs underneath (T-45).
This threat is close to rendered-vs-signed divergence (T-7) but sits in a different family: in T-7 the TS legitimately belongs to the user and the fault is that the signed fields diverge from the rendered summary, so the corruption is of the mandate \emph{content} (Semantic Manipulation); in T-45 the consent surface itself is hijacked, so the user's approval action is captured for a mandate they never saw, and the corruption is of the \emph{authority} the signature is attributed to. In all cases, AP2 accepts an action as authoritative without being able to prove that the right authority actually produced or approved it.

\end{description}

\subsection{F3: Supply-Chain \& Trust-Root Subversion}
\label{subsec:f3-provenance-subversion}

These attacks force AP2 to trust or run illegitimate dependencies before the mandate checks can protect the transaction. This family contains 13 of the 48 threats (T-18--T-30). They primarily target AG1 and AG2, with AG4 and AG6 as further consequences.

\begin{description}
\item[{Software Provenance.}]
This sub-family covers cases where an AP2 role runs, loads, or trusts code, models, prompts or tool infrastructure that the operator did not intend to deploy. It includes AP2 build or CI/CD compromise (T-18), non-attestable builds that make swapped artifacts hard to detect (T-19), runtime compromise of an agentic role (T-20), malicious MCP server images (T-21), unverified LLM model substitution (T-22), and sub-agent prompt drift across releases (T-23). The common failure is provenance: AP2 relies on software or model state whose origin and integrity are not reliably bound to the transaction.

\item[{Discovery Trust.}]
The SA establishes trust with an illegitimate peer or registry during discovery, before any AP2 artifacts are generated. This includes AgentCard integrity failures on both the publish and consume side, since AgentCards are unsigned by spec and can be tampered at origin, in transit, or in cache (T-24) \cmdd{}; discovery transport MITM or redirection via a rogue root CA, host network misconfiguration or DNS/well-known hijack (T-25); a malicious registry or registry caching that subverts the trust root itself (T-26); and an A2A AP2-extension downgrade where a look-alike or fail-open extension mismatch silently drops AP2 enforcement (T-27). 

\item[{Channel Identity.}]
Trust is broken at the channel or tool-identity layer mid-session, after discovery. This includes an MCP server identity that is not session-bound, so a mid-session redirect to a malicious MCP breaks no AP2 invariant (T-28); cross-server tool or resource confusion through typosquatted tool names or overlapping resource handles that route data to the wrong server (T-29); and missing message-level role authentication, where post-TLS AP2 channels carry no app-layer role signature, leaving handshakes, status replies, and side-channel fields role-spoofable (T-30).
\end{description}

\subsection{F4: State-Binding Failures}
\label{subsec:f4-state-binding}

State-binding failures occur when AP2 fails to maintain a secure and consistent transaction state across its operational phases. Their goals include AG3 as primary and AG1, AG6, AG4 and sometimes AG5, depending on whether the attacker reuses a valid state, creates an inconsistent state or destabilizes historical verification. This family contains 9 of the 48 threats (T-31--T-38, T-46).

\begin{description}
\item[{Execution-State Binding.}]
These attacks break the relationship between the conditions the user authorized and the state actually consumed during execution. This includes replaying a closed mandate or fanning out one open mandate across multiple executions (T-31); mutating cart or constraint state after user review but before signing (T-32); shared-MCP races or ordering attacks that leave peers with conflicting transaction state (T-33); shared-MCP cache bleed across peers without concurrency (T-46); and unsigned \texttt{risk\_data} fields that bias CP/MPP step-up or fraud decisions (T-34).

\item[{Historical Binding.}]
These attacks break cryptographic re-verification over extended periods. This includes linkability from weak SD-JWT salts (T-35); loss of historical verifiability when key rotation or missing trust snapshots prevent reconstruction of the signing-time trust state (T-36); stale mandate reuse, where AP2 does not tie a receipt to the validity window of the mandate that authorized it, so a stale receipt can settle a lapsed authorization (T-37); and long-term cryptographic aging, where retained ECDSA-P256 mandates may outlive their security time frame (T-38). The common failure is that AP2 cannot reliably bind a past transaction to the trust and cryptographic state that existed when it occurred.

\end{description}

\subsection{F5: Accountability Failures}
\label{subsec:f5-accountability}

Accountability failures occur when AP2 validates a proxy identity or keeps an incomplete audit trail, allowing the actual entity responsible for execution, fees, or liability to remain hidden. Their primary goal is AG5 (Repudiation); depending on the hidden principal's action, they also realize AG4 and AG1. This family contains 7 of the 48 threats (T-39--T-44, T-47).

\begin{description}
\item[{Dispute-Evidence Gaps.}]
These attacks exploit structural gaps in what AP2 can prove after the fact. This includes missing evidence that AP2 cannot prove ever existed (T-39); absent per-role, per-decision or consent-ceremony audit trails (T-40); missing behavioral telemetry that lets compromise resemble normal traffic (T-41); deletion requirements that destroy dispute evidence (T-44); and MCP elicitation prompts that mimic AP2 consent without producing a TS audit trail (T-47). The dispute resolver is left with a partial evidence chain and cannot identify who saw, approved, changed, or retained the relevant state.

\item[{Marketplace Fronting.}]

While the marketplace identity is visible to AP2, the sub-merchant that actually shaped the transaction remains hidden.
A single marketplace \texttt{merchant\_identity} can hide the sub-merchant or intermediary responsible for transaction terms, fees, fulfillment, dispute facts, or post-signing behavior (T-43). The signed cryptographic chain points exclusively to the front identity, even though the accountable principal sits behind it.

\item[{Multi-Tenant Isolation.}]
Here a CP or MPP serves many tenants from one platform instance but fails to isolate their state. Session context, cached decisions, or scoped tokens issued for one tenant bleed into another, either exposing that tenant's payment data or biasing an authorization or step-up decision made on its behalf. Because every request is signed and processed under the shared platform's identity, the AP2 mandate chain attributes the action only to that platform service and cannot name the tenant whose state actually shaped the outcome (T-42).
The failure is architectural rather than cryptographic: AP2 binds signatures to platform identities, not to the tenants, sub-merchants, and evidence chains that actually shape and account for the transaction

\end{description}
\section{Attack Demonstrations and Mitigations}
\label{sec:attack-demonstrations}
This section presents five attack demonstrations and mitigations (\texttt{AM1}--\texttt{AM5}) that together cover all eight high risk threats identified in Figure~\ref{fig:threat-matrix}. The first two are chains, in which each link creates a precondition the next needs or disables a defense that would otherwise stop it, so the attack fails without any one link; the remaining three each realize a single high risk threat in isolation.
Each demonstration describes the attack scenario, attacker requirements, access and knowledge capabilities, deployment architecture, the threats realized, attack families and attacker goals, and the mitigations.

\tcbset{
    attackbox/.style={
        enhanced jigsaw,
        breakable,
        width=\linewidth,
        before skip=3pt,
        after skip=3pt,
        boxsep=1.2pt,
        top=1.5pt,
        bottom=1.5pt,
        left=2pt,
        right=2pt,
        sharp corners,
        boxrule=0.45pt,
        colback=white,
        colframe=black!70,
        colbacktitle=black!75,
        coltitle=white,
        fonttitle=\bfseries\attackfontsize,
        fontupper=\attackfontsize,
        toptitle=1pt,
        bottomtitle=1pt,
        lefttitle=3pt,
        righttitle=3pt,
        before upper={
            \setlength{\parskip}{0pt}
            \setlength{\parindent}{0pt}
            \renewcommand{\arraystretch}{1}
        }
    }
}

\begin{tcolorbox}[attackbox, title={AM1. Authorization Beyond Stated Intent (\texttt{T-23} $\rightarrow$ \texttt{T-1} $\rightarrow$ \texttt{T-4})}]
A malicious merchant wants the SA to obtain broader payment authority than the user intended. The user asks the SA to buy one camera for no more than \$50. The deployment uses HNP mode and a multi-agent SA whose consent sub-agent compares proposed open mandates with the user's request. Its original rule requires escalation whenever a proposed amount exceeds the user's limit or a requested restriction is absent. A prompt-only release changes that rule so that the sub-agent accepts deviations when the surrounding transaction context appears to justify them (P1, \texttt{T-23}). The merchant may exploit drift already present in the deployment.

The merchant then supplies catalog terms claiming that a \$50 purchase may require up to \$80 of authorization headroom and that the payee will be resolved at checkout (P2 and P3, \texttt{T-1}). Its MCP quote tool returns an amount range from \$50 to \$80 but no payee field. The SA maps that result into a \texttt{payment.amount\_range} constraint with \texttt{min=5000}, \texttt{max=8000}, and \texttt{currency=USD}. Because the tool supplies no payee, the mapper emits no \texttt{payment.allowed\_payees} constraint (P3, \texttt{T-4}). The drifted consent sub-agent treats the merchant-controlled terms and quote as sufficient justification for both deviations and does not escalate.

The TS presents the open mandates for approval. The review identifies the task as buying one camera expected to cost \$50 and presents the \$80 ceiling and unrestricted payee status as checkout headroom. Relying on the consent workflow's lack of a warning and the merchant's explanation, the user authorizes the open mandates. This scenario does not require the TS to hide signed fields. If the TS displayed only the \$50 purchase while signing the broader constraints, the deployment would also realize rendered-versus-signed divergence (\texttt{T-7}).

The merchant later creates a merchant-signed checkout for the same camera at \$80, and the SA signs the corresponding closed Checkout and Payment Mandates with its bound agent key. 

\par\vspace{1pt}
\begin{tabularx}{\linewidth}{@{}>{\bfseries}lX@{}}
Requirements. & HNP mode and a prompt-defined consent decision. \\
Access. & \texttt{ACR4} on \texttt{S2}; \texttt{TA2} catalog-authoring influence on \texttt{S11}; \texttt{ACR5} on \texttt{S9}. \\
Knowledge. & \texttt{AK1}, \texttt{AK2}, \texttt{AK3}, \texttt{AK4}. \\
Architecture. & A3$\times$A2 in HNP mode; amplified under A5. \\
Threats. & \texttt{T-23}, \texttt{T-1}, \texttt{T-4}. \\
Family / goals. & \texttt{F3}/\texttt{T-23}; \texttt{F1}/\texttt{T-1}, \texttt{T-4}. \texttt{AG2}; \texttt{AG1} if the \$80 checkout settles. \\
\end{tabularx}

\par\vspace{1pt}
\textbf{Mitigation.}
For \texttt{T-23}, sub-agent prompts, thresholds, and tool grants become versioned release artifacts, with a provenance record for every change. The SA rejects a prompt identity that is not tied to an approved release, and a deterministic versioned policy outside the model enforces the escalation rule.
For \texttt{T-1}, catalog terms and tool results retain their origin and remain transaction data rather than authorization. Merchant content may report a price or fee, but it cannot justify increasing a ceiling or removing a restriction.
For \texttt{T-4}, the deployment pins the tool schema and passes results through a typed mapper. Before signing, the TS checks and explicitly renders the exact value of \texttt{payment.amount\_range.max}, rather than a generic amount range, and states when \texttt{payment.allowed\_payees} is absent. It rejects a maximum above the user's limit or an absent payee restriction that the user did not explicitly authorize, and signs only the bytes it rendered.
\end{tcolorbox}

\begin{tcolorbox}[attackbox, title={AM2: Role Confusion at a Shared Tool Layer (\texttt{T-29} $\rightarrow$ \texttt{T-15})}]
\textbf{Attack Scenario.}
A camera store and its largest competitor sell through the same payments platform, where the shopping agents, merchant agents, and the Credentials Provider all reach their tools through one shared MCP layer (a generalized A5 instance, Section~\ref{subsec:ap2-deployment-arch}). The competitor runs one of the tool servers on that shared layer, ordinary for a merchant peer under \texttt{TA2}, and uses it to make the CP act on its behalf. The opening is that AP2 authenticates artifacts but never checks who is behind a tool call: which server answered it, and which role made it.

It starts with a name collision. The competitor registers \texttt{checkout.quote\_v2} under a name matching the canonical \texttt{checkout.quote}, and because the shopping agent resolves tool names against every connected server with nothing tying a name to the one server allowed to answer it, some quote requests reach the competitor's server. It returns correct quotes, so nothing looks wrong, but every request reveals the fields the shopping agent attached: the live transaction and cart identifiers, the amounts, and the CP handle in use (\texttt{T-29}).

That context enables the second step. Holding a real transaction identifier, the competitor calls the CP's credential tools and binds them to it. The CP authorizes on the transport channel the call arrives on, since AP2 carries role identity in signed artifacts and not in tool calls, and a merchant peer's channel looks legitimate; unable to tell the caller is reaching for CP-reserved operations, it answers (\texttt{T-15}). Neither step suffices alone: the collision only leaks context and confers no authority, while a role-confused call with no live transaction identifier binds to nothing that matters. Together they let a merchant peer drive CP-reserved operations, and because every call is signed under the platform's identity, the audit trail names only the platform, never the server that answered or the role that called.

\par\vspace{1pt}
\begin{tabularx}{\linewidth}{@{}>{\bfseries}lX@{}}
Requirements. & Tool names must resolve across servers without a binding to an authoritative server identity, and the shared component must authorize by transport rather than by AP2 role. \\
Access. & \texttt{ACR5} on \texttt{S9} to operate the shadowing tool server, and \texttt{ACR1} on the shared \texttt{S9}. \\
Knowledge. & \texttt{AK1}, \texttt{AK2}. \\
Architecture. & Generalized A5: one MCP tool layer shared across the SA, MA, and CP. \\
Threats. & \texttt{T-29}, \texttt{T-15}. \\
Family / goals. & \texttt{F3}/\texttt{T-29}; \texttt{F2}/\texttt{T-15}. \texttt{AG1}. \\
\end{tabularx}

\par\vspace{1pt}
\textbf{Mitigation.}
Both steps close by binding an identity that AP2 currently leaves to the channel: the server that answers, and the role that calls.
For \texttt{T-29}, the SA pins each tool to the one server allowed to answer it and rejects name collisions across servers. The binding is fixed at session start and cannot be re-resolved mid-session, so a look-alike tool never receives the call.
For \texttt{T-15}, every MCP call carries an application-layer signature over caller identity, AP2 role, and transaction identifier, and the CP authorizes on that signature rather than on the transport channel. CP-reserved tools are further gated by per-caller capability tokens issued for the transaction, so a legitimate peer channel alone never confers a role.
\end{tcolorbox}

\begin{tcolorbox}[attackbox, title={AM3: Cross-Tenant Credential Theft (\texttt{T-42})}]
\textbf{Attack Scenario.}
A payments platform runs one CP instance for all the merchant programs it hosts, and a small merchant on that platform is the attacker. It has a legitimate storefront, a legitimate AP2 role, and, like every tenant, a legitimate channel to the CP. Its goal is a rival store's customer payment credentials, the tokens the CP mints when it approves a checkout.

The weakness is in how the CP remembers approvals. It issues each payment credential as a scoped token and files it in one shared store keyed by the token and the transaction reference, with no tenant in the key. The attacker learns this indexing is flat: nothing in a lookup says which merchant a reference belongs to. So during its own ordinary checkouts it harvests transaction references, then replays them to the CP's credential interface under its own valid channel. The CP resolves each reference against the shared store, finds the matching approval, and returns the token, without ever asking whether the caller is the merchant that reference was created for. One of those tokens belongs to the rival's customer, and the CP hands it over. Because the call is signed under the platform's own identity and the store carries no tenant, the audit trail records only that the platform issued a credential, never that a competitor pulled a rival customer's token from shared state.

\par\vspace{1pt}
\begin{tabularx}{\linewidth}{@{}>{\bfseries}lX@{}}
Requirements. & The CP must serve multiple tenants without per-tenant cryptographic scoping of session, cache, and token state. \\
Access. & \texttt{ACR1} on \texttt{S5} at the verifier interface. Isolation being absent, the standalone path assumed \texttt{ACR2} is not required. \\
Knowledge. & \texttt{AK1}. \\
Architecture. & A4: a multi-tenant CP verifier. \\
Threats. & \texttt{T-42}. \\
Family / goals. & \texttt{F5}. \texttt{AG4} for the cross-tenant disclosure; \texttt{AG3} where the bled decision inflates the released scope. \\
\end{tabularx}

\par\vspace{1pt}
\textbf{Mitigation.}
Every entry in the token and decision store carries the tenant it belongs to, and the CP admits a lookup only when the caller's authenticated tenant matches that entry, so a token or approval filed for one merchant is never returned to another. The token itself is derived under a per-tenant key, and the responsible tenant, not only the platform, is named in the audit trail.
\end{tcolorbox}

\begin{tcolorbox}[attackbox, title={AM4: AP2 Extension URI Downgrade (\texttt{T-27})}]
\textbf{Attack Scenario.}
A shopper's SA begins a purchase, and an attacker positioned on the discovery path answers first. At \texttt{P2}, before any AP2 artifact exists to be signed, the attacker returns an unverified A2A AgentCard advertising a Merchant endpoint whose AP2 extension URI differs from the canonical one by a single character: \url{https://registry.ap2-protoco1.org/v2/secure} against \url{https://registry.ap2-protocol.org/v2/secure}, the digit "1" standing in for the letter "l". The SA matches this URI loosely rather than exact-matching it against a pinned canonical value, so the look-alike passes and the SA binds the attacker's endpoint to the session. The trap is that this endpoint speaks only AP2 v0.1: the transaction silently downgrades and loses the context-binding and replay defenses v0.2 introduced~\cite{lan_2026} (\texttt{T-27}). The whole substitution happens before artifact validation begins, so every later signature check still passes over a session that is already replayable and unbound, and neither the shopper nor the merchant sees a downgrade occur.

\par\vspace{1pt}
\begin{tabularx}{\linewidth}{@{}>{\bfseries}lX@{}}
Requirements. & The SA must match the extension URI loosely rather than exact-match it against a pinned value, and accept the downgraded discovery result before AP2 artifact validation begins. \\
Access. & \texttt{ACC2} or \texttt{ACR1} on \texttt{S8}. \\
Knowledge. & \texttt{AK1}. \\
Architecture. & A1 and above; the attack depends on no MCP, multi-agent, or marketplace surface. \\
Threats. & \texttt{T-27}. \\
Family / goals. & \texttt{F3}. \texttt{AG1}, \texttt{AG2}. \\
\end{tabularx}

\par\vspace{1pt}
\textbf{Mitigation.}
During \texttt{P2}, the SA exact-matches the extension URI against a pinned canonical value or a signed AgentCard rather than matching loosely. A URI that does not match exactly makes the SA fail closed: the legacy fallback is refused and the transaction dropped rather than silently downgraded.
\end{tcolorbox}

\begin{tcolorbox}[attackbox, title={AM5: Parameter Poisoning in \texttt{risk\_data} (\texttt{T-34})}]
\textbf{Attack Scenario.}
A fraudster holds a stolen card number and wants to spend it without triggering the step-up challenge that would ask for something only the real cardholder has. Running its own SA, it drives the transaction itself, and the SA controls the one input the verifiers trust for that decision: the \texttt{risk\_data} field carried with the Payment Mandate. In the AP2 reference implementation~\cite{ap2spec}, the SA populates \texttt{risk\_data} as a mutable field forwarded with the mandate, and no verifier checks its origin or integrity, so the fraudster simply writes the values it wants: a low risk score and a step-up-completed claim that never happened. The CP reads the supplied score and releases the payment credential as low risk; the MPP reads the step-up-completed flag and skips the OTP or 3DS2 challenge, the exact control meant to stop a stolen-card charge. Where the deployment forwards the same \texttt{risk\_data} to a payment-network risk engine, the planted low-risk signal further lowers the chance the charge is ever flagged as fraud.

\par\vspace{1pt}
\begin{tabularx}{\linewidth}{@{}>{\bfseries}lX@{}}
Requirements. & The deployment must accept SA-authored or A2A-carried \texttt{risk\_data} as verifier input. \\
Access. & \texttt{ACR2}/\texttt{ACR3} on the user's own \texttt{S2}, with write influence over \texttt{risk\_data} carried with \texttt{S10}. \\
Knowledge. & \texttt{AK1}, \texttt{AK4}. \\
Architecture. & A1 and above. \\
Threats. & \texttt{T-34}. \\
Family / goals. & \texttt{F4}. \texttt{AG1}; \texttt{AG3} where the suppressed step-up widens the effective authorization. \\
\end{tabularx}

\par\vspace{1pt}
\textbf{Mitigation.}
During \texttt{P3}, \texttt{risk\_data} is accepted only inside the signed Payment Mandate or as a signed attestation the mandate references, conforming to a registered schema and issued by an authorized source such as the TS. The SA may not attach it as a mutable field. At \texttt{P4}, the CP and MPP reject any transaction whose risk payload is unsigned, off-schema, or unattributable to an authorized source.
\end{tcolorbox}

\section{The AP2 Security Scanner Tool}
\label{sec:scanner}

We built an AP2 security scanner to evaluate implementations against the threat catalog using static source analysis and targeted runtime checks. Given an implementation's source code, the scanner reports evidence-backed findings mapped to concrete attack paths across its roles, tools, verifier logic, and protocol surfaces. The scanner is available in the companion repository.\footnote{\url{https://anonymous.4open.science/r/AP2_Beyond_the_Mandate}}
The scanner has four stages. First, it profiles the deployment's architecture, active AP2 roles, agentic roles, and exposed surfaces. Second, it filters the threat taxonomy matrix to threats whose architecture and surface requirements match that profile. Third, it runs threat-specific checks tied to each retained threat's surface and failure mode. For example, it checks whether MCP tools enforce per-request authorization through role-bound capability tokens rather than shared caller state, and whether AP2 control fields come only from typed policy or quote objects rather than open schema boundaries.
Some failures require cross-role analysis. The Cross-Role Differential Specification Evaluator (CDSE) compares the constraint schemas, mandate fields, and accepted tool schemas that the SA, MA, CP, and MPP use for the same transaction. It flags cases in which one role signs, verifies, or executes a semantic object that differs from the object used by another role. This detects Semantic Manipulation failures in which a mandate is cryptographically valid but the transaction details differ across roles or diverge from the user's intent.
Implementations that run in a test harness can also use an optional adversarial mode. The scanner starts a temporary test instance, reuses the discovered role wiring and tool definitions, selects a user-defined top-$K$ set of threats with testable runtime paths, and runs fixed, reproducible probes against those paths. It records a finding only when a probe meets its threat-specific success condition.
Finally, the report records each finding's evidence, affected role, surface, phase, threat ID, attack family, and required mitigation. This lets implementers identify exposed threats and determine whether the corresponding mitigations are enforced in code or at runtime.
\providecommand{\hit}{$\bullet$}
\providecommand{\uniq}{$\bullet^{*}$} 

\section{Evaluation}
\label{sec:evaluation}

We evaluate the threat catalog, the AIVSS risk assessment, and the scanner's layered design through three research questions.

\begin{description}
    \item[\textbf{RQ1. Catalog comparison.}]
    To what extent does a structured STRIDE-GPT analysis elicit the threats in the MAESTRO-derived catalog?

    \item[\textbf{RQ2. Risk-score reproducibility.}]
    To what extent do independent experts reproduce the AIVSS ratings when applying the CVSS and AARS rubric defined in Section~\ref{subsec:aivss-risk-assessment}?

    \item[\textbf{RQ3. Scanner-layer contribution.}]
    Which threats does each scanner layer detect that the other layers miss?
\end{description}

\subsection{Threat Catalog Comparison (RQ1)}
\label{subsec:eval-stride}

\paragraph{Method}
We used STRIDE-GPT~\cite{stridegpt} as an independent threat-modeling baseline. STRIDE~\cite{shostack2014} examines components and data flows, whereas MAESTRO examines agents and their interactions. This comparison tests which catalog threats a structured STRIDE analysis elicits and whether it finds any threat absent from our catalog. We considered three agent-specific alternatives. ATFAA provides an agent-focused threat taxonomy but no automated comparison procedure~\cite{atfaa2025}. STRIDE-AI and ASTRIDE were relevant, but we found no public artifacts that allowed us to reproduce their analyses~\cite{strideai,astride}. We excluded the official MAESTRO Analyzer because it uses the framework from which we built our catalog and would not provide an independent baseline~\cite{maestro_analyzer}.

We used STRIDE-GPT v0.18.0 and ran it once for each deployment architecture from A1 through A5. Both configured generation roles used \texttt{gpt-5.3}. Each input described the same protocol roles, HP/HNP cryptographic artifacts, security properties, and AP2 lifecycle. Only the architecture block changed. The inputs excluded MAESTRO terminology, catalog identifiers and titles, target mechanisms, examples, expected counts, and prior mappings. The catalog and STRIDE-GPT runs used the same lifecycle and security-property definitions, which may increase measured agreement. One run per architecture also does not measure run-to-run variation.

An LLM judge compared every catalog threat with all STRIDE-GPT rows. Each row represented a generated threat scenario. The judge processed one catalog item per call and used \texttt{gpt-5.5} at temperature $0$ with a fixed seed, so the generator did not grade its own output. For each item, the judge proposed a verdict, supporting-row identifiers, and a rationale. The mapping was many-to-many. Several rows could jointly support one catalog item, and one row could support several items. \textsc{Full} required compatible rows to cover the affected component or artifact, action or failure, security effect, and relevant architecture conditions. \textsc{Partial} covered the core failure and security effect but missed or generalized at least one AP2-specific detail without contradicting the item. \textsc{None} meant that no row, alone or in combination, could derive the threat. We manually checked every proposed \textsc{Full} and \textsc{Partial} mapping against the evidence and rules. We did not include \textsc{None} labels in that review. We also checked every STRIDE-GPT row for a catalog mapping. An unmapped row would be a candidate addition.

Two third-party annotators independently assessed every \textsc{None} item. Each had five years of cybersecurity experience and had developed multi-agent applications. They received the threat catalog, the AP2 documentation supplied to the judge, and an XLSX form requesting a label, confidence rating, and notes. \emph{Distinct} meant that the item was a valid AP2 threat and did not duplicate a positively matched item. \emph{Overlap} meant that a positively matched item already covered substantially the same threat. \emph{Unsupported} meant that the AP2 system model did not support the threat.

\textit{Results}
The five STRIDE-GPT runs produced 124 rows, with 24 each for A1, A2, A3, and A5 and 28 for A4. The judge labeled 32 of the 48 catalog items \textsc{Full} or \textsc{Partial} and the remaining 16 \textsc{None}. We agreed with 31 of the 32 judgments and changed one from \textsc{Full} to \textsc{Partial}. Full-or-partial coverage was $32/48=66.7\%$. All 124 STRIDE-GPT rows mapped to at least one catalog item, so none qualified as a candidate addition.

For the 16 \textsc{None} items, Annotator~1 assigned 15 \emph{Distinct} labels and one \emph{Overlap} label. Annotator~2 assigned all 16 items \emph{Distinct}. They agreed on 15 items, giving raw agreement of $15/16=93.75\%$. They resolved the remaining disagreement as \emph{Distinct}. The final labels were 16 \emph{Distinct}, zero \emph{Overlap}, and zero \emph{Unsupported}.

\begin{table}[t]
\centering
\caption{Catalog coverage from five frozen structured STRIDE-GPT runs. }
\label{tab:stride-coverage}
\footnotesize
\setlength{\tabcolsep}{2.5pt}
\begin{tabular}{@{}lrrrr@{}}
\toprule
Family & Full & Partial & Not elicited & Total \\
\midrule
\rowcolor{red!8}    F1 Semantic                  & 4 & 2 & 4 & 10 \\
\rowcolor{orange!10}F2 Authority                 & 5 & 3 & 1 & 9 \\
\rowcolor{yellow!12}F3 Supply-chain/trust-root   & 5 & 4 & 4 & 13 \\
\rowcolor{blue!8}   F4 State binding             & 3 & 3 & 3 & 9 \\
\rowcolor{green!10} F5 Accountability            & 2 & 1 & 4 & 7 \\
\midrule
Total & \textbf{19} & \textbf{13} & \textbf{16} & \textbf{48} \\
\bottomrule
\end{tabular}
\end{table}

\subsection{Risk Assessment Reproducibility (RQ2)}
\label{subsec:eval-irr}

\paragraph{Method}

One rater assigned the AIVSS scores reported in Section~\ref{subsec:aivss-risk-assessment}. Eight independent participants from academia and industry then applied the same rubric. We tested whether the bands derived from their ratings reproduced the original severity bands. Our primary hypothesis was that the lower bound of the 95\% confidence interval for Gwet's AC2 would be at least $0.61$. For each threat, raters completed the 11 CVSS v4.0 base metrics~\cite{cvss40} and ten AARS factors~\cite{aivss}. They did not calculate an AIVSS score or select a band. We reconstructed each score and band using the procedure in Section~\ref{subsec:aivss-risk-assessment}. Disagreement therefore reflects metric and factor assignments rather than arithmetic or band selection.

Each rater first completed two calibration examples, which we excluded from the analysis. The instrument then presented 25 threats. It included five threats per family, spanned the represented bands, and included threats near the High threshold. For each threat, the questionnaire fixed the architecture with the highest band. Five threats (\texttt{T-1}, \texttt{T-15}, \texttt{T-21}, \texttt{T-34}, \texttt{T-42}) formed a common set rated by all participants. We split the remaining 20 into two non-overlapping batches of ten, each covering all five families. Every participant rated the five common threats and one batch. Four participants rated Batch~A and four rated Batch~B. Each group included two participants from academia and two from industry. They worked independently, and we pseudonymized their responses.

We measured agreement with quadratically weighted Gwet's AC2~\cite{gwet2008}. This coefficient gives partial credit to adjacent-band disagreements and is less sensitive than $\kappa$ to skewed frequencies. We cross-checked it with quadratically weighted Krippendorff's $\alpha$~\cite{krippendorff2004} and exact pairwise agreement. The AC2 interval uses a leave-one-threat-out jackknife over all 25 threats. For $\alpha$, we resample the 25 threats with replacement and report the middle 95\% of the resulting estimates. The $0.61$ hypothesis threshold is an operational reference drawn from the Landis and Koch interpretation of $\kappa$~\cite{landis1977}. It is not a validated cutoff for quadratically weighted AC2, so we interpret AC2 together with the two complementary agreement measures.

\textit{Results}
Table~\ref{tab:irr-agreement} reports agreement on the derived severity bands.

\begin{table}[t]
\centering
\caption{Inter-rater agreement on AIVSS severity bands, reported as
estimates with 95\% confidence intervals.}
\label{tab:irr-agreement}
\scriptsize
\setlength{\tabcolsep}{2.5pt}
\renewcommand{\arraystretch}{1.2}
\begin{tabular}{@{}>{\raggedright\arraybackslash}p{0.28\columnwidth}
                   >{\raggedright\arraybackslash}p{0.15\columnwidth}
                   >{\raggedright\arraybackslash}p{0.26\columnwidth}
                   >{\raggedright\arraybackslash}p{0.21\columnwidth}@{}}
\toprule
Statistic & Scope & Est.\ [95\% CI] & Interpretation \\
\midrule
\rowcolor{gray!14}
\textbf{Gwet's AC2 (quad.)} & all 25 &
$\mathbf{0.984}$ {\scriptsize[0.963,\,1.000]} &
LB exceeds 0.61 benchmark \\

Krippendorff's $\alpha$ (quad.) & all 25 &
$0.802$ {\scriptsize[0.520,\,0.978]} &
High estimate; CI crosses 0.61 \\

Exact pairwise agreement & all 25 &
$0.923$ {\scriptsize[0.846,\,0.980]} &
High observed agreement; no chance correction \\
\bottomrule
\end{tabular}
\end{table}

Quadratic weighting gives substantial credit to adjacent-band disagreements. With that qualification, the AC2 lower bound exceeds the $0.61$ reference point. Exact pairwise agreement and Krippendorff's $\alpha$ provide complementary evidence. Participants also reproduced the single-rater reference band in 111 of 120 ratings ($92.5\%$).

\subsection{Scanner-Layer Ablation (RQ3)}
\label{subsec:eval-ablation}

\paragraph{Method}

The scanner in Section~\ref{sec:scanner} has three layers. The deterministic \emph{Passive} layer checks source code and configuration. The LLM-driven \emph{CDSE} layer compares specifications across AP2 roles. The LLM-driven \emph{Adversarial} layer runs threat-specific probes against a test instance. We used an ablation study to measure each layer's contribution.

No public AP2 deployment was available, so we used our AP2 testbed. Section~\ref{sec:current-limit} discusses this limitation. Recall is the fraction of applicable threats detected for each architecture. The evaluation includes five threats for A2, five for A3, two for A4, and six for A5. A threat counts as detected when any enabled layer reports it. Each finding records the layer that produced it.

We evaluated seven configurations. \textbf{Full} enables all three layers. Three leave-one-out configurations disable one layer (\textbf{No-Passive}, \textbf{No-CDSE}, and \textbf{No-Adv}), and three isolated configurations enable one (\textbf{Only-Passive}, \textbf{Only-CDSE}, and \textbf{Only-Adversarial}). We ran every configuration five times per architecture. Every run that invoked CDSE or Adversarial used OpenAI \texttt{gpt-5.5}. We provide the scanner prompts in the GitHub repository.

The Adversarial layer tests only threats with an executable Bounded Adversarial Sandbox strategy. Each strategy defines an attack procedure and success condition for a shadow agent configured with the target's system prompt and tool schemas. Eligible threats must also have an architecture-specific AIVSS score of at least $4.0$. If more than eight qualify, the scanner tests the eight highest-ranked threats. None of the evaluated architectures exceeded this limit.
We classify a detection as \emph{unique} when one isolated layer reports the threat in all five runs and the other two report it in none.

\paragraph{Results}

The Full configuration attained the highest mean recall on every architecture, although other configurations tied it on A4 and A5 (Table~\ref{tab:abl-recall}).

\begin{table}[t]
\centering
\caption{Recall of the seven scanner configurations, reported as
mean $\pm$ population standard deviation across five runs.}
\label{tab:abl-recall}
\footnotesize
\setlength{\tabcolsep}{4pt}
\resizebox{\columnwidth}{!}{%
\begin{tabular}{@{}l>{\columncolor{gray!10}}ccccccc@{}}
\toprule
Arch & Full
& \shortstack{No-\\Pass.}
& \shortstack{No-\\CDSE}
& \shortstack{No-\\Adv}
& \shortstack{Only-\\Pass.}
& \shortstack{Only-\\CDSE}
& \shortstack{Only-\\Adv} \\
\midrule
A2 & .80
   & $.36{\scriptstyle\pm.08}$
   & .80
   & .60
   & .60
   & .00
   & $.36{\scriptstyle\pm.08}$ \\

A3 & .60
   & $.28{\scriptstyle\pm.10}$
   & .40
   & .60
   & .40
   & .20
   & $.08{\scriptstyle\pm.10}$ \\

A4 & .50
   & .50
   & .00
   & .50
   & .00
   & .50
   & .00 \\

A5 & 1.00
   & $.93{\scriptstyle\pm.08}$
   & .83
   & 1.00
   & .83
   & $.87{\scriptstyle\pm.16}$
   & .17 \\
\bottomrule
\end{tabular}%
}
\end{table}

The isolated configurations showed which detections each layer added. Passive uniquely detected \texttt{T-4}, \texttt{T-8}, and \texttt{T-24}. CDSE uniquely detected \texttt{T-14}, \texttt{T-1} on A4, and \texttt{T-29}. Adversarial uniquely detected \texttt{T-5}. Each layer detected at least one seeded threat that both other layers missed, so no layer subsumed the others. All three layers missed \texttt{T-2} on A2, \texttt{T-1} and \texttt{T-9} on A3, and \texttt{T-3} on A4. Table~\ref{tab:abl-matrix} gives the full per-threat attribution.

\begin{table}[t]
\centering
\caption{Per-threat detection by the isolated scanner layers. \hit\ denotes detection in all five runs. A fraction gives the number of detections across five runs. \uniq\ denotes detection in all five runs by one layer and none by the other two.}
\label{tab:abl-matrix}
\footnotesize
\setlength{\tabcolsep}{5pt}
\begin{tabular}{@{}lccc@{}}
\toprule
Threat & Passive & CDSE & Adversarial \\
\midrule
\multicolumn{4}{@{}l}{\emph{A2, Single-Agent with Isolated MCP}}\\
T-2  & 0/5   & 0/5   & 0/5   \\
T-4  & \uniq & 0/5   & 0/5   \\
T-8  & \uniq & 0/5   & 0/5   \\
T-34 & \hit  & 0/5   & 4/5   \\
T-5  & 0/5   & 0/5   & \uniq \\
\addlinespace
\multicolumn{4}{@{}l}{\emph{A3, Multi-Agent}}\\
T-1  & 0/5   & 0/5   & 0/5   \\
T-9  & 0/5   & 0/5   & 0/5   \\
T-24 & \uniq & 0/5   & 0/5   \\
T-27 & \hit  & 0/5   & 2/5   \\
T-14 & 0/5   & \uniq & 0/5   \\
\addlinespace
\multicolumn{4}{@{}l}{\emph{A4, Marketplace}}\\
T-3  & 0/5   & 0/5   & 0/5   \\
T-1  & 0/5   & \uniq & 0/5   \\
\addlinespace
\multicolumn{4}{@{}l}{\emph{A5, Shared MCP}}\\
T-1  & \hit  & \hit  & 0/5   \\
T-15 & \hit  & \hit  & 0/5   \\
T-31 & \hit  & \hit  & 0/5   \\
T-33 & \hit  & 3/5   & \hit  \\
T-46 & \hit  & 3/5   & 0/5   \\
T-29 & 0/5   & \uniq & 0/5   \\
\bottomrule
\end{tabular}
\end{table}

\section{Discussion and Future Work}

\subsection{Discussion}
\label{sec:discussion}

The threat modeling, attack taxonomy, and PoC validation yield three observations relevant to AP2 deployments.

\subsubsection{Agentic authorization must bind operational context, not only signed intent}
\label{subsec:disc-context-binding}

Across our PoCs, AP2 artifacts remain cryptographically valid even when a transaction no longer reflects the user's intent. Signatures cover the mandate but may omit the operational context from which the agent constructs it, including discovery choices, intermediate cart state, tool outputs, and verifier-facing disclosures. When an LLM-driven role selects this context, the pre-signing path becomes an attack surface.

Several PoCs succeed without forging signatures or breaking cryptographic checks. They steer the agent before it produces a protected AP2 artifact, so mandate-chain verification may not reveal the failure. The same limitation affects Tamarin or ProVerif analyses when their models cover signed artifacts but omit the agentic execution path. Agentic authorization protocols should bind relevant pre-signature context to issued artifacts or reject execution paths that do not produce such bindings.

\subsubsection{AP2 security depends on its protocols and deployment architecture}
\label{subsec:disc-substrate-architecture}

AP2 defines the authorization layer but leaves A2A transport and MCP tool-access architecture to implementers. AP2 consumes data from these protocols before committing it to a signed mandate, and many high-severity threats in our catalog originate in those exchanges. We therefore examined how A2A and MCP vulnerabilities affect mandate semantics and verifier decisions. This dependency is asymmetric: AP2 mitigations can validate the final mandate but cannot retroactively secure the flawed protocol exchange that produced it. Several mitigations must therefore be implemented as structural requirements in the supporting protocols.

The same asymmetry applies to deployment architecture. AP2 does not prescribe how A2A endpoints, MCP servers, agents, and merchant components are deployed. These choices determine the capabilities an attacker gains from a compromise and the trust boundaries between roles. As a result, a threat may be absent in one architecture, present in another, and amplified in a third. Deployments using the same protocols can therefore face different threats.

\subsubsection{Different threat classes require different validation techniques}
\label{subsec:disc-driveability}

The scanner maps each threat class to a suitable validation strategy. Reasoning-layer vulnerabilities require adversarial testing, whereas supply-chain, cryptographic, and data-retention flaws are better suited to static analysis and configuration review. Because no single method covers the full attack surface, agentic-protocol security tools must combine detection methods matched to specific threat classes.

\subsection{Limitations}
\label{sec:current-limit}

\subsubsection{No complete public AP2 deployment}

At the time of evaluation, no complete public production or prototype AP2 deployment was available; Google's public AP2 implementation covers only part of the protocol~\cite{ap2spec}. We therefore built the AP2 testbed used for our PoCs. This limits our ability to claim that the same failures will appear unchanged in future third-party deployments (T-34). However, the testbed allowed us to evaluate AP2 before large-scale adoption, while design and implementation guidance can still influence how the protocol is deployed. The testbed, threat catalog, and scanner can help implementers identify trust boundaries, attack surfaces, and failure modes before production release.
Our PoCs show that the attacks can affect authorization within AP2, including attacks that originate in A2A or MCP interactions. They do not measure whether or how the resulting AP2-layer failures affect card networks, issuers, acquirers, or other payment-rail systems.

\subsection{Future Work}
\label{sec:future-work}

Once public AP2 deployments or reference implementations become available, future work should run the scanner against them to audit the threats identified here. Such evaluations would provide a more realistic basis for estimating false-positive and false-negative rates and help turn the scanner from a research prototype into a production-ready tool.
\section{Conclusion}
\label{sec:conclusion}

AP2 introduces a cryptographic mandate chain that makes user intent verifiable, binds checkout and payment mandates, and yields artifacts verifiable after a transaction. In this work, we analyze how unsecured catalog data, tool results, A2A messages or some other context information can manipulate agents before signing. A manipulated pre-signing path can yield a valid mandate chain authorizing a transaction the user did not intend. We therefore analyzed AP2 end to end as an agentic payment system.
We decomposed AP2~v0.2 into five lifecycle phases and five deployment architectures; built a MAESTRO-based threat model covering four threat actors, eleven attack surfaces, and six attacker goals; and grouped 48 threats into five attack families. AIVSS rated 8 threats high in at least one architecture. We also built a testbed for the PoC demonstrations and developed a scanner to operationalize the security analysis. As AP2 moves from specification to production, the threat catalog and scanner can help implementers identify architecture-specific threats and enforce the required mitigations.

\bibliographystyle{ACM-Reference-Format}
\bibliography{references}

\end{document}